\documentclass[twocolumn]{aa} 

\usepackage{graphicx}
\usepackage{txfonts}
\usepackage{mathabx}

\usepackage{comment}
\usepackage{ulem}
\usepackage{lineno}
\usepackage{color}
\graphicspath{{./}{./figures/}}

\begin{document} 

\title{Monitoring the radio emission of Proxima Centauri}
\titlerunning{Monitoring the radio emission of Proxima Centauri}

\author{M. P\'erez-Torres \inst{1}
          \and J.F. G\'omez\inst{1}
          \and J.L. Ortiz\inst{1}
          \and P. Leto\inst{2}
          \and G. Anglada\inst{1}
          \and J.L. G\'omez\inst{1}
          \and E. Rodr\'iguez\inst{1}
          \and C. Trigilio\inst{2}
          \and P.J. Amado\inst{1}
          \and A. Alberdi\inst{1}
          \and G. Anglada-Escud\'e\inst{3}
          \and M. Osorio\inst{1}
          \and G. Umana\inst{2}
          \and Z. Berdi\~nas\inst{4}
          \and M.J. L\'opez-Gonz\'alez\inst{1}
          \and N. Morales\inst{1}
          \and C. Rodr\'iguez-L\'opez\inst{1}
          \and J. Chibueze\inst{5,6}
          }

\institute{CSIC, Instituto de Astrof\'isica de Andaluc\'ia, 
     Glorieta de la Astronom\'ia S/N, 
E-18008, Granada, Spain\\ \email{torres@iaa.es}
         \and
         INAF, Osservatorio Astrofisico di Catania, Via S. Sofia 78, 
I-95123 Catania, Italy
         \and
        School of Physics and Astronomy, Queen Mary University of London, 327 Mile End
        Road, London E1 4NS, UK
        \and
        Departamento de Astronom\'ia, Universidad de Chile, 
Camino El Observatorio, 1515 Las Condes, Santiago, Chile
        \and
        Centre for Space Research, Potchefstroom campus, North-West University,
Potchefstroom 2531, South Africa
        \and
        Department of Physics and Astronomy, Faculty of Physical Sciences,  University of Nigeria, Carver Building, 1 University Road,
Nsukka, Nigeria
             }

\date{Received MM DD, YYYY; accepted December 2, 2020}

\abstract{We present results from the most comprehensive radio monitoring
    campaign towards the closest star to our Sun, Proxima Centauri. We report
    1.1 to 3.1 GHz observations with the Australian Telescope Compact Array
    over 18 consecutive days in April 2017.  We detect radio emission from
    Proxima Centauri for most of the observing sessions, which spanned
    $\sim$1.6 orbital periods of the planet Proxima b.  The radio emission is
    stronger at the low-frequency band, centered around 1.6 GHz, and is
    consistent with the expected electron-cyclotron frequency for the known
    star's magnetic field intensity of $\sim$600 Gauss.  The 1.6 GHz light
    curve shows an emission pattern that is consistent with the orbital period
    of the planet Proxima b around the star Proxima, with its maxima of
    emission happening near the quadratures.  We also observed two
    short-duration (a few minutes) flares and a long-duration (about three
    days) burst whose peaks happened close to the quadratures.  We find that
    the frequency, large degree of circular polarization, change of the sign of
    circular polarization, and intensity of the observed radio emission are all
    consistent with expectations from electron cyclotron-maser emission arising
    from sub-Alfv\'enic star-planet interaction.  We interpret our radio
    observations as signatures of interaction between the planet Proxima b and
    its host star Proxima.  We advocate for monitoring other dwarf stars with
    planets to eventually reveal periodic radio emission due to star-planet
interaction,  thus opening a new avenue for exoplanet hunting and the study of
a new field of exoplanet-star plasma interaction.}


\maketitle
%

\section{Introduction} \label{sec:intro}

The finding of a planet in the habitable zone of the dwarf M star Proxima
Centauri (hereafter Proxima) has represented a major breakthrough in
exoplanetary science, especially because the mass and size of the planet is
likely similar to that of Earth \citep{ang+16}. This discovery has triggered
plenty of renewed interest in our close neighbor Proxima, which has been
subject to many observational campaigns at different wavelengths of the
electromagnetic spectrum, including the detection at mm wavelengths of thermal
emission from the star itself, and possibly from circumstellar material
\citep{ang+17}.

\cite{lim+96} and \cite{sle+03} reported the detection of radio emission
towards Proxima at wavelengths of $\sim$20 cm with the Australian Telescope
Compact Array (ATCA).  More recently, \cite{bel+16} reported the non-detection
of radio emission from Proxima at 154 MHz, using 12 observations with the
Murchison Widefield Array spread between 10 February 2014 and 30 April 2016,
placing a 3-$\sigma$ upper limit on the steady-state radio emission from the
system in Stokes I of 42.3 mJy\,beam$^{-1}$.

The emission detected by \cite{lim+96} and \cite{sle+03} presented a degree of
circular polarization of nearly 100\%, and those authors discussed several
possible mechanisms, including electron cyclotron-maser (ECM) mechanism, but
did not provide any details on the origin of the observed emission.  The ECM
mechanism is also responsible for the highly polarized periodic radio pulses
detected in different classes of stars, ranging from hot B/A-type magnetic
stars \citep{tri+00,das+18,das+19a,das+19b,let+19,let+20} to ultra-cool dwarfs
and brown dwarfs \citep{hal+07,ber+09,rou+12,kao+16,zic+19}. 

After the discovery of the exoplanet Proxima Centauri b (hereafter Proxima b)
with an orbital period of 11.186 days \citep{ang+16}, we considered the
feasibility of carrying out a monitoring campaign to study Proxima and Proxima
b in detail at radio wavelengths.  The detection of direct radio emission from
the planet Proxima b itself would be very difficult from ground-based radio
telescopes because the known emission mechanisms would result in emission
either with a very low frequency (below the low-frequency end of the radio
window set by the Earth's ionosphere), or very weak (see, e.g.,
\citealt{zar07}; \citealt{kat+16}). Nevertheless, the interaction of Proxima
and its planet Proxima b could produce detectable radio emission at centimeter
wavelengths.  \cite{zar07} reviewed several possible plasma interactions of
exoplanets with their parent stars, and their associated radio emission.  In
particular, star-planet interaction can produce ECM emission in a way
equivalent to the Jupiter/Io interaction, with the star playing the role of
Jupiter, and the exoplanet that of Io.  The characteristic frequency of the ECM
emission is given by the electron gyrofrequency, $\nu_g = 2.8\,B_\ast$ MHz,
where $B_\ast$ is the stellar surface magnetic field, in Gauss. For Jupiter,
which has a weak magnetic field ($\sim$4.2 G at the equator;
\citealt{zar+96,con+18}), the gyrofrequency falls in the decametric range.  In
contrast, the magnetic field in the surface of Proxima  is $B_\ast = 600 \pm
150$ G \citep{rei+08}, so if the ECM mechanism is at place in the Proxima
system as a result of star-planet interaction, emission at a frequency of
$\nu_g \simeq$ 1.7 $\pm$ 0.4 GHz should be expected.

\section{Observations and data processing} \label{sec:obs}

\noindent We observed Proxima with the Australia Telescope Compact Array (ATCA)
in 2017 April 12-29, at a central frequency of 2.1 GHz (corresponding to a
wavelength of $14.3$ cm).  Our observations consisted of 18 observing daily
epochs between 2017 April 12-29, encompassing 1.5 orbital cycles (the orbital
period of Proxima b is of 11.2 days).  Each observing session lasted for 3
hours, except the one on 2017 April 24, which was 12 hour-long to obtain a full
synthesis map of the whole field of view. For all observing sessions, the ATCA
was in its 6A configuration, which yields maximum baselines of 5938.8 m. We
recorded data using the Compact Array Broadband Backend in CFB\_1M mode, which
allowed us to observe over a bandwidth of 2 GHz centered at 2.1 GHz, sampled
over 2048 channels of 1 MHz width each, and the basic integration time was of
10 s.  The system yielded auto- and complex cross-correlation products of two
perpendicular, linearly polarized signals, from which we obtained full
polarization products (all 4 Stokes parameters).  We used the source PKS
1934-638 as bandpass and absolute flux calibrator in all sessions except on
April 22, when it was not visible from ATCA, so we used PKS 0823-500 instead.
For complex gain calibration, we used the source PKS 1329-665 on April 12, and
PMN J1355-6326 in the rest of sessions. The typical duty cycle for the
observations lasted for $\simeq 19$ min, with 15 min on target (Proxima), 3 min
on the complex gain calibrator, and the remaining time spent on antenna
slewing. For all observations, we set the phase center at R.A.
$=14^h29^m33.456^s$, Dec$=-62^\circ40'32.89''$ (J2000.0).

We used the  Miriad package \citep{sau+95} for data editing and calibration.
After applying calibration, we exported the visibilities as a measurement set,
and performed all imaging steps within the Common Astronomy Software
Applications (CASA) package \citep{mcm+07}. All images presented in this paper
were obtained by using multifrequency synthesis and Brigg's weighting (with
robust parameter 0.5 as defined in CASA) on the visibility data,  and were
deconvolved with the CLEAN algorithm.  The resulting synthesized beams of the
images covering the whole 2 GHz-wide band were $4.4''\times 3.8''$ on April 24
(when we had a 12-hr full observing track on Proxima), and $\simeq 20''\times
4''$ for the rest of the observing epochs. We obtained maps of total (Stokes I)
and circularly polarized flux density (Stokes V = RCP $-$ LCP, where RCP and
LCP are right and left circular polarization, respectively) for each day, and
in different frequency ranges.  We determined the flux densities of Proxima
from the peak intensity of each image within one beam of the source position,
using task \texttt{imfit} of CASA, and considered the source as detected when
that peak was above three times the rms of the map and its location was
consistent with the expected position of Proxima.  Positional uncertainties are
estimated to be $\simeq 1''-4''$, considering errors  in the absolute
astrometry of ATCA at the observed frequency band \citep{cas98} and in the fit
of the peak position  (which are of the order of the half-width synthesized
beam divided by the signal-to-noise ratio; \citealt{con+98}). Given the
frequency dependence of the emission discussed below, we obtained images at two
different frequency ranges: one of 400 MHz bandwidth, centered at 1.62 GHz
($18.5$ cm), and another one of 1 GHz bandwidth, centered at 2.52 GHz ($11.9$
cm).

We searched for variability on short timescales by analyzing the
interferometric visibilities, using both the task DFTPL of the \textsl{AIPS}
package and the task \texttt{uvmodelfit} of the \textsl{CASA} software package.
DFTPL plots the discrete Fourier transform (DFT) of the complex visibilities
for any arbitrary point in the sky, as a function of time. This task is useful
to study the time variability of an unresolved source without the need of
making a synthesis map for each time interval. To isolate the emission of
Proxima, we made a map of the whole field of view for each observing epoch, and
deconvolved it with the CLEAN algorithm, except a tight region around Proxima.
Using the resulting CLEAN components as a model of the background sources, we
run the task \texttt{uvsub} of \textsl{CASA} to subtract the model from the
visibilities of the corresponding day. Finally, we run DFTPL on the resulting
visibilities using different time intervals.  As a trade-off between
signal-to-noise and temporal resolution, we examined the data over intervals of
a few times 10 sec, which is the duration of a scan.  As an independent test,
we also run the task \texttt{uvmodelfit} of \textsl{CASA} on the
background-subtracted visibilities for analogous time intervals. This
\textsl{CASA} task fits a point source to the visibilities. The results
obtained with DFTPL and \texttt{uvmodelfit} are consistent within the
uncertainties, and below we discuss only the former.

\section{Results} \label{sec:results}

\subsection{Long-term radio variability and its correlation with the orbital phase of the planet} 
\label{sec:images}

\begin{figure}[thbp]
  \centerline{\includegraphics[width=\linewidth]{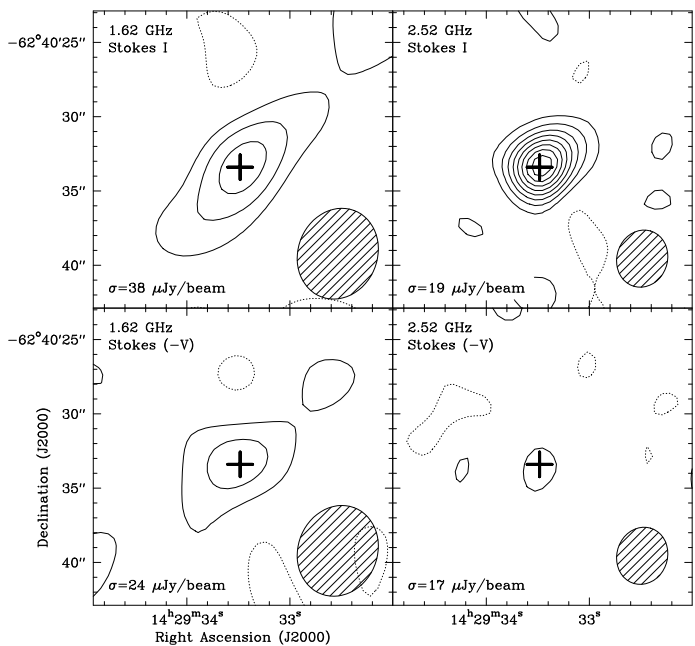}}
  \caption{\label{fig:proxima_IVmaps}
  Contour maps of radio continuum emission from the Proxima Centauri
system on 24 April 2017. 
The maps correspond to total (Stokes I) and circularly
polarized (Stokes V) flux density, for two bands: one of 400 MHz width
centered at 1.62 GHz, and another one of 1 GHz width centered at 2.52
GHz. 
Contour levels are drawn at $-$2, 2, 4, 6, 8, 10, 12, 14, and 16 times the rms of each
map (denoted by $\sigma$ in each panel). 
We inverted the sign of the Stokes V maps for better
visualization, so solid contours in the two bottom panels actually
represent negative values of the circularly polarized flux
density. The hatched ellipse at the bottom right corner of each panel
represents the half power contour of the synthesized
beams. The cross corresponds to the position of the star Proxima for the
observing epoch, obtained from the 1.3 mm ALMA data reported in
\cite{ang+17}, taking into account the proper motion and
parallax of the star.}
\end{figure}

\begin{figure}[thbp]
    \centerline{\includegraphics[width=0.9\linewidth]{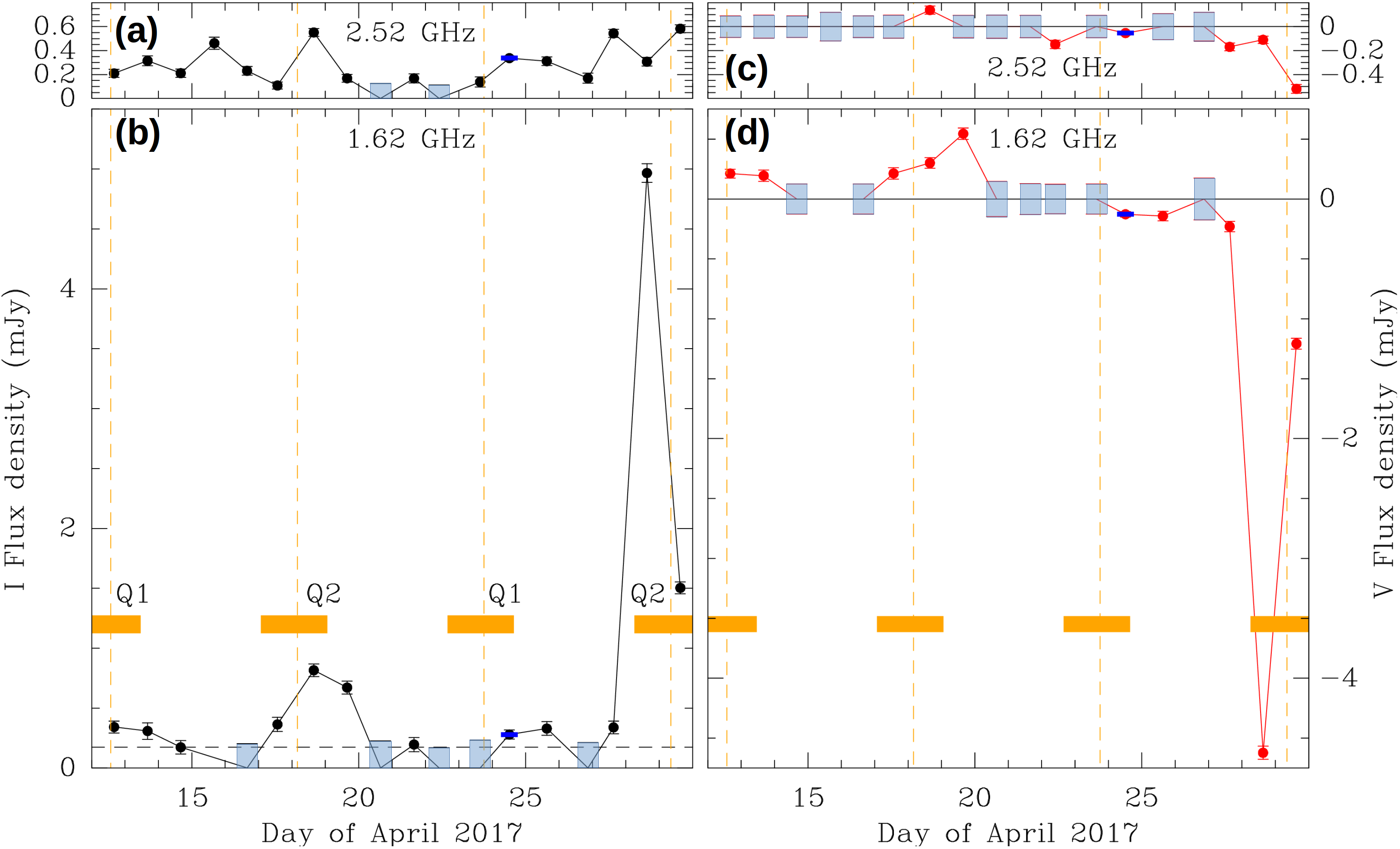}}
    \vspace{0.3cm}
    \centerline{\includegraphics[width=0.9\linewidth]{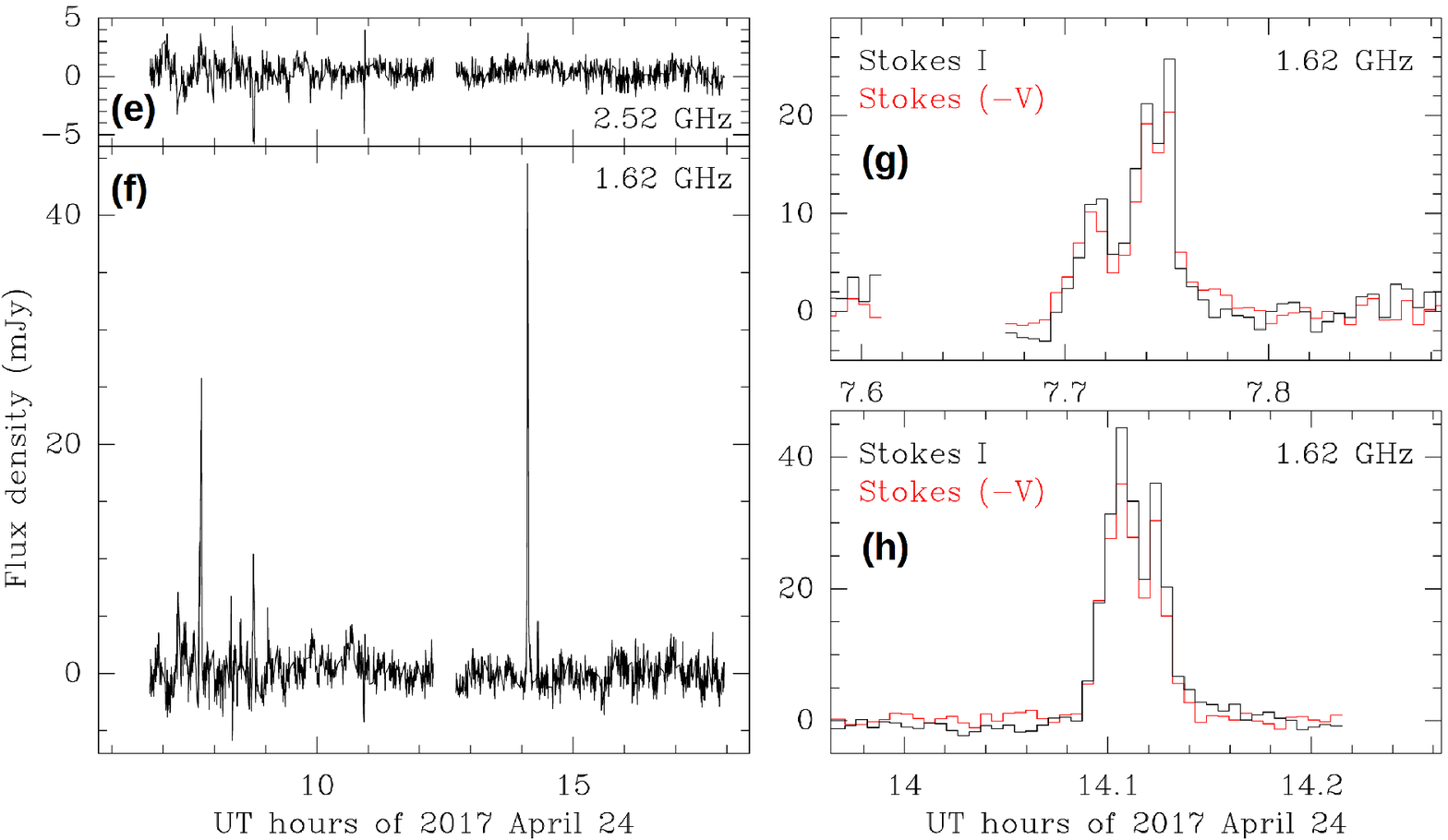}}
    \caption{\label{fig:flux_evolution} Time evolution of flux density in
        Proxima. Variation of total (Stokes I; panels \textbf{a} and
        \textbf{b}) and circularly polarized (Stokes V; panels \textbf{c} and
        \textbf{d}) flux density as a function of time during our ATCA
        observing campaign. For each observing session, we averaged the data
        over a bandwidth of 400 MHz and 1 GHz centered at 1.62 GHz  [wavelength
        $\simeq 18.5$ cm; \textbf{(b,d)}] and 2.52 GHz [$\simeq 11.9$ cm;
        \textbf{(a,c)}], respectively.  1$\sigma$ uncertainties are plotted as
        vertical lines for each data point, and blue shaded rectangles mark
        3$\sigma$ upper limits for non-detections (in the case of Stokes V,
        they correspond to upper limits to the absolute value of the flux
        density).  The duration of the observation on April 24 (12 hours) is
        represented with a horizonal blue bar, while the observing time of the
        other days ($\sim$2 hours) is smaller than the symbol size of data
        points.  Dashed orange lines show the quadratures, Q1 and Q2, of
        Proxima b, with the horizontal orange bars indicating the uncertainty
        in the determination of the epoch of the quadratures (see main text for
        details).   The horizontal dashed line in the plot of Stokes I at 1.62
        GHz \textbf{(b)} corresponds to a flux density of $0.174\pm 0.038$ mJy,
        obtained by averaging together the data over the five observing
        sessions where the source was not detected individually at that
        frequency. No map could be obtained at low frequency on April 15, due
        to an insufficient number of unflagged visibilities.  \textbf{(e)-(f)}
        Variation of Stokes I during 2017 April 24 for data averaged over 20 s
        intervals. Two short-duration flares  are evident at 1.62 GHz
        \textbf{(f)}. \textbf{(g)-(h)} Temporal close-up of the two flares.
        Black and red lines correspond to Stokes I and V, respectively. The
        sign of Stokes V has been reversed for better visualization.  }
    \end{figure}

Radio emission from the Proxima system is detected in most observing epochs
(Table \ref{tab:data}).  As an illustration, in Fig. \ref{fig:proxima_IVmaps}
we show the maps of the radio continuum emission over the two frequency bands
(centered at 1.62 and 2.52 GHz, corresponding to wavelengths of 18.5 and 11.9
cm, respectively), on 2017 April 24. The data taken that day had the best
interferometric coverage of all our observational campaign, as this session
lasted for 12 hr.  The location of the emission is consistent, within
astrometric uncertainties, with the estimated position of Proxima at the epoch
of the observations \citep{ang+17}.  The total flux density (Stokes I) over the
whole 2-GHz wide bandwidth, was 0.304 $\pm $ 0.017 mJy on April 24, which
corresponds to a monochromatic radio luminosity of (6.14$\pm$0.34)$\times
10^{11}$ erg\,s$^{-1}$\,Hz$^{-1}$.  Circularly polarized flux density (Stokes
V) was also detected on that day, with Proxima being the only source in the
field that showed circular polarization above the noise level.  The maps in
Fig. \ref{fig:proxima_IVmaps} exclude two short-duration flares, each of
$\sim$4 min duration (Fig.  \ref{fig:flux_evolution}f-h), as the variable
emission from those flares produces spurious features in the images, when
included.  The measured flux density, however, was very similar in images with
and without those short-duration flares (see Sect. \ref{sec:short-duration}).   

\begin{figure}[thbp]
    \centerline{\includegraphics[width=\linewidth]{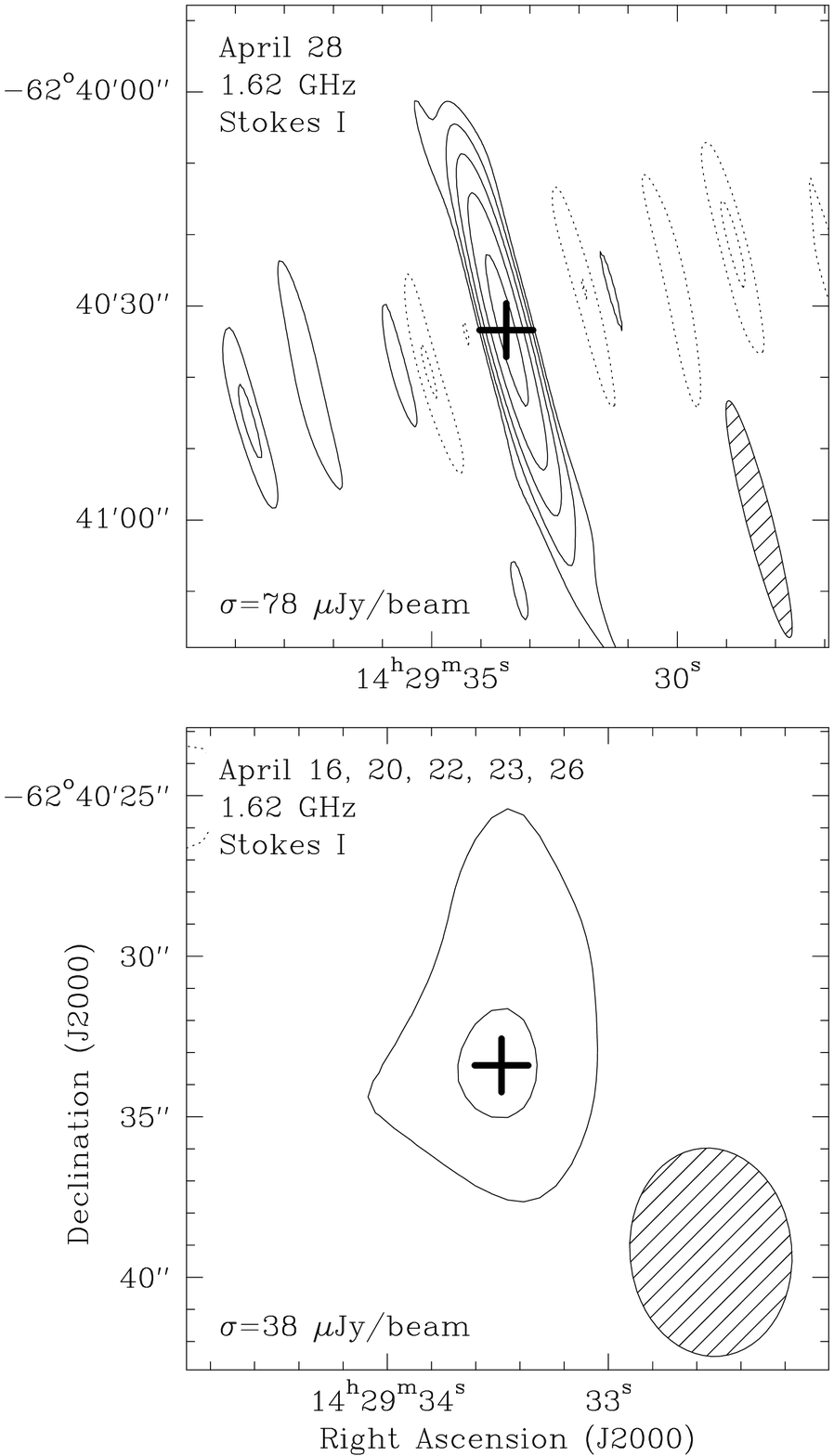}}
    \caption{\label{fig:proxima_minmax_maps} Contour maps of  Proxima at 1.62
        GHz during maximum and minimum emission. Top panel: Emission on 2017
        April 28, when the flux density was at its maximum.  Contour levels are
        drawn at $-$6, $-$3, 3, 6, 12, 24, and 48 $\times$ 78 $\mu$Jy
        beam$^{-1}$, the rms of the map.  Bottom panel: Map obtained by
        combining the uv data for the observing epochs when the emission was
        not detected in each of the individual images (2017 April 16, 20, 22,
        23, and 26).  Contour levels are drawn at -2, 2, and $4\,\times 38\,
        \mu$Jy beam$^{-1}$, the rms of the map. In both maps, solid and dotted
        contours represent positive  and negative levels, respectively.  }
    \end{figure}

In Fig.~\ref{fig:flux_evolution}a-d and Table \ref{tab:data}, we show the
evolution of the Stokes I and Stokes V data towards the Proxima system,
averaged over each observing session, and in the two frequency bands.  The
Stokes I flux density of the low-frequency band, centered at 1.62 GHz
($\Delta\nu = $400 MHz),  has an average value of $\simeq 0.31$ mJy over the
whole observing period.  This value corresponds to an in-band isotropic  radio
power $P_{r} \approx 2.51\times10^{20}$ erg s$^{-1}$, for an assumed solid
angle $\Omega = 4\,\pi$ sr.  The radio emission of Proxima shows  significant
variability over the observing campaign, especially at the low-frequency band.
Indeed, the 1.62 GHz Stokes I flux density clearly shows several increases over
the quiescent state, each lasting 2-3 days, with an especially long burst
during the last 3 days of the observations, which reached $\simeq$5 mJy.  This
flux density corresponds to a brightness temperature $T_{\rm b} \gtrsim
3.1\times10^{11}\,[\Delta l/ (0.1 R_\ast)]^{-2}$ K, where $R_\ast =
0.145\,R_\odot$ is the radius of the Proxima Cen star, and $\Delta l$ is the
size of the emitting region, which we have normalized to the typical size of a
stellar magnetic loop \citep{lop+06}.  In Fig.  \ref{fig:proxima_minmax_maps}
we show Stokes I images at 1.62 GHz, on April 28 (the day with the highest flux
density at this frequency), and for the combined data of April 16, 20, 22, 23,
and 26.  While the source is not detected above the 3$\sigma$ threshold in any
of the individual epochs of the combined image, it clearly shows emission at a
level of 0.174 $\pm $ 0.038 mJy, indicating the presence of a relatively weak,
yet quiescent radio emission from Proxima.

\begin{table*}
\caption{\label{tab:data}\textbf{Log from our ATCA observations}}
\begin{minipage}{\textwidth}
\begin{center}
\footnotesize
\begin{tabular}{lrrrrrrrr}
\multicolumn{9}{c}{\normalsize Low-frequency (1.62 GHz) observations}\\
\hline
     Time\footnote{Mean time of each observing session, given in days of April
     2017} &   Phase\footnote{Orbital phase of Proxima b, measured from 0 to 1}
     &    Phase\footnote{Orbital phase of Proxima b, measured in degrees from 0
     to 360}  &  Stokes I\footnote{Total flux density of Proxima. For
     non-detections, 3$\sigma$ upper limits are given} &  rms(I)\footnote{Rms
     of total flux density} & Stokes V\footnote{Circularly polarized flux
     density. For non detections, we give 3$\sigma$ upper limits to its absolute value.}
                            & rms(V)\footnote{Rms of circularly polarized flux density.}
                            & $p_V$\footnote{Fraction of circular polarization (V/I).
   When I is detected, but not V, we give upper limits to the polarization fraction as
   $3\times\mbox{rms(V)}/{\rm I}$.} & $\sigma(p_V)\footnote{1-$\sigma$ uncertainty in
 the fraction of circular polarization, $p_V$.}$
     \\
   (days) &   (0-1) &  (degrees) &       (mJy) &  (mJy/beam) &       (mJy) &  (mJy/beam) \\
\hline
12.673 &  0.2593 &      93.34 &       0.342 &  0.049 &       0.212 &  0.037 & 0.62 & 0.14\\
13.677 &  0.3490 &     125.65 &       0.309 &  0.069 &       0.194 &  0.048 & 0.63 & 0.21 \\
14.669 &  0.4377 &     157.58 &       0.173 &  0.056 &    $\mid <$0.126$\mid$ &  0.042 & $<0.73$\\
15.675\footnote{For the epoch 15.675 April, the low-frequency data had to be severely
flagged, so obtaining an image was not possible.} &      -- &         -- &          -- &     -- &          -- &     -- &   -- &  -- \\
16.656 &  0.6154 &     221.53 &      $<$0.204 &  0.068 &    $\mid <$0.126$\mid$ &  0.042 \\
17.567 &  0.6968 &     250.85 &       0.364 &  0.059 &       0.213 &  0.047 & 0.59 & 0.16\\
18.654 &  0.7940 &     285.84 &       0.816 &  0.053 &       0.301 &  0.044 & 0.37 & 0.06\\
19.654 &  0.8834 &     318.01 &       0.672 &  0.053 &       0.545 &  0.047 & 0.81 & 0.09\\
20.657 &  0.9731 &     350.32 &      $<$0.228 &  0.076 &    $\mid <$0.147$\mid$ &  0.049 \\
21.656 &  0.0624 &      22.45 &       0.196 &  0.059 &    $\mid <$0.129$\mid$ &  0.043 & $<0.66$\\
22.407 &  0.1295 &      46.61 &      $<$0.171 &  0.057 &    $\mid <$0.123$\mid$ &  0.041 \\
23.637 &  0.2394 &      86.20 &      $<$0.234 &  0.078 &    $\mid <$0.126$\mid$ &  0.042 \\
24.514 &  0.3179 &     114.44 &       0.279 &  0.038 &      -0.127 &  0.024 &0.46 & 0.11\\
25.634 &  0.4180 &     150.48 &       0.330 &  0.057 &      -0.142 &  0.040 & 0.43 \\
26.867 &  0.5281 &     190.13 &      $<$0.213 &  0.071 &    $\mid <$0.174$\mid$ &  0.058 \\
27.634 &  0.5968 &     214.84 &       0.339 &  0.053 &      -0.230 &  0.043 &0.68 & 0.17\\
28.634 &  0.6862 &     247.02 &       4.967 &  0.078 &      -4.623 &  0.055 &0.93 & 0.02\\
29.634 &  0.7756 &     279.22 &       1.504 &  0.050 &      -1.208 &  0.045 &0.80 & 0.04\\
\hline \\
\multicolumn{9}{c}{\normalsize High-frequency (2.52 GHz) observations}\\
\hline
12.673 &  0.2593 &      93.34 &       0.210 &  0.034 &    $\mid <$0.090$\mid$ &  0.030 & $<0.43$\\
13.677 &  0.3490 &     125.65 &       0.315 &  0.041 &    $\mid <$0.096$\mid$ &  0.032 & $<0.30$\\
14.669 &  0.4377 &     157.58 &       0.212 &  0.035 &    $\mid <$0.090$\mid$ &  0.030 & $<0.43$\\
15.675 &  0.5277 &     189.96 &       0.462 &  0.050 &    $\mid <$0.120$\mid$ &  0.040 & $<0.26$\\
16.656 &  0.6154 &     221.53 &       0.231 &  0.036 &    $\mid <$0.090$\mid$ &  0.030 & $<0.39$\\
17.567 &  0.6968 &     250.85 &       0.108 &  0.030 &    $\mid <$0.096$\mid$ &  0.032 & $<0.89$\\
18.654 &  0.7940 &     285.84 &       0.551 &  0.033 &       0.139 &  0.034 & 0.25 & 0.06\\
19.654 &  0.8834 &     318.01 &       0.167 &  0.033 &    $\mid <$0.093$\mid$ &  0.031 & $<0.55$\\
20.657 &  0.9731 &     350.32 &      $<$0.126 &  0.042 &    $\mid <$0.096$\mid$ &  0.032 \\
21.656 &  0.0624 &      22.45 &       0.168 &  0.038 &    $\mid <$0.093$\mid$ &  0.031 & $<0.55$\\
22.407 &  0.1295 &      46.61 &      $<$0.126 &  0.042 &      -0.148 &  0.034 \\
23.637 &  0.2394 &      86.20 &       0.139 &  0.042 &    $\mid <$0.093$\mid$ &  0.031 & $<0.67$\\
24.514 &  0.3179 &     114.44 &       0.337 &  0.019 &      -0.053 &  0.017 &0.16 & 0.05\\
25.634 &  0.4180 &     150.48 &       0.312 &  0.036 &    $\mid <$0.108$\mid$ &  0.036 & $<0.35$\\
26.867 &  0.5281 &     190.13 &       0.169 &  0.042 &    $\mid <$0.120$\mid$ &  0.040 & $<0.71$\\
27.634 &  0.5968 &     214.84 &       0.544 &  0.034 &      -0.167 &  0.032 & 0.31 & 0.06\\
28.634 &  0.6862 &     247.02 &       0.307 &  0.036 &      -0.110 &  0.030 & 0.36 & 0.11\\
29.634 &  0.7756 &     279.22 &       0.583 &  0.034 &      -0.520 &  0.037 &0.89 & 0.08\\
\hline
\end{tabular}
\end{center}
\end{minipage}
\end{table*}

The low-frequency (1.62 GHz) data is strongly circularly polarized (typically
40\%--80\%), reaching 80\%--90\% during the long burst
(Fig.~\ref{fig:flux_evolution} and Table \ref{tab:data}).  The radio emission
shows also a remarkable inversion of its circular polarization, with the Stokes
V $>0$ for the first half of our observations (until around April 20), and
V$<0$ from April 24 onwards (panels b and d of Fig.~\ref{fig:flux_evolution}).
The high degree of circularly polarized emission indicates that the mechanism
responsible for the observed emission is coherent.

\begin{figure}[thbp] 
  \centerline{\includegraphics[width=\linewidth]{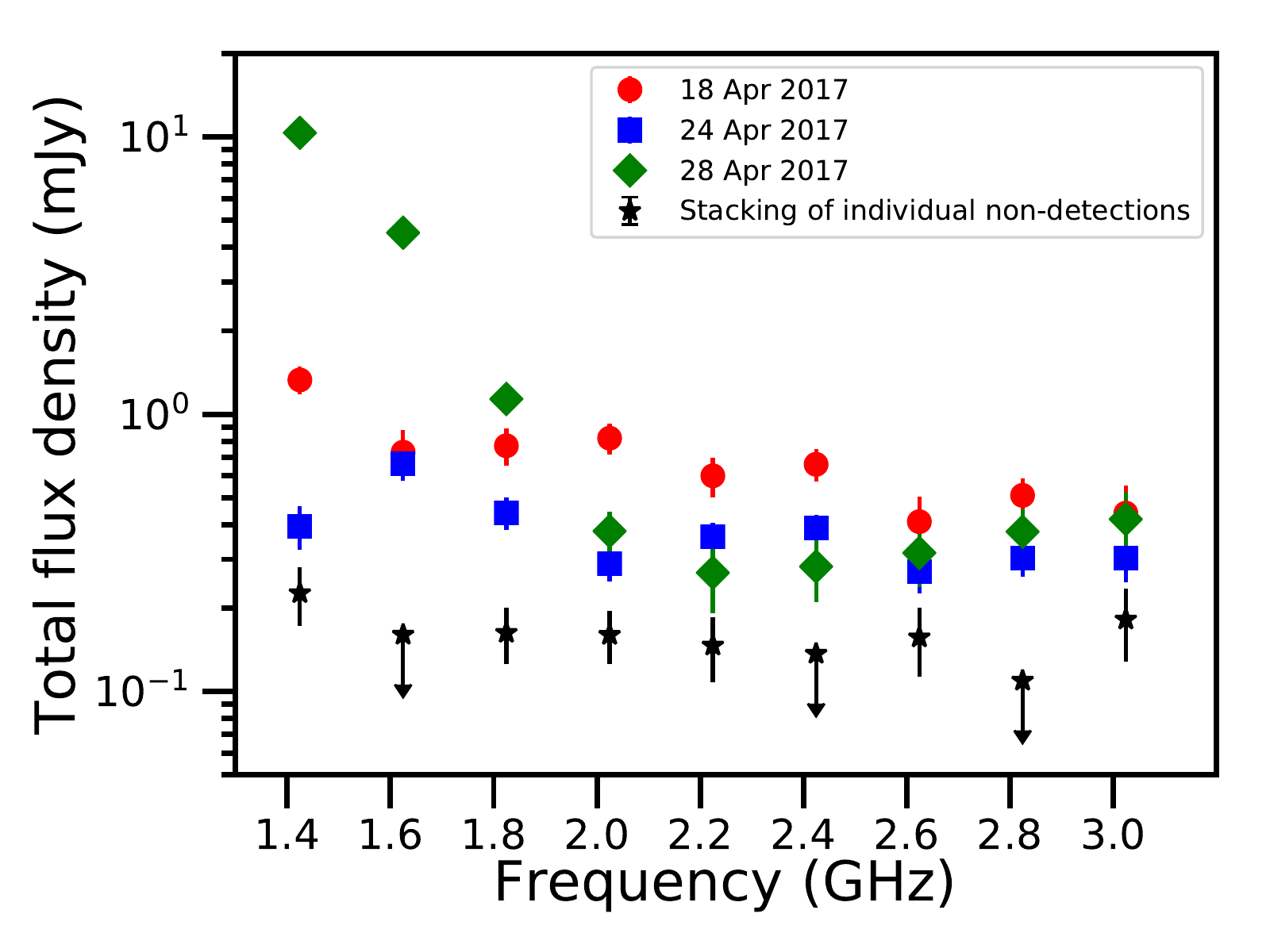}}
  \caption{\label{fig:Stokes-I-vs-freq}
    Illustration of the day-to-day radio spectral evolution of Proxima. 
    Data correspond to the total flux density (Stokes I) on 2017 April 18 (circles), April 24
    (squares), and April 28 (diamonds) over the whole observing bandwidth,
    averaged every 200 MHz.  For comparison, we also show the result of stacking the
    data corresponding to the five individual observing epochs where there was no
    detection (data of April 16, 20, 22, 23, and 26), drawn as stars. Arrows indicate 3-$\sigma$ upper limits.
}
\end{figure}

The variability and degree of circular polarization are significantly lower at
the higher frequency band (2.52 GHz; Fig.~\ref{fig:flux_evolution} and Table
\ref{tab:data}). This is also illustrated in Fig.~\ref{fig:Stokes-I-vs-freq},
where we show the ATCA radio spectrum of Proxima over the observing bandwidth
(from 1.3 up to 3.1 GHz) for three representative days: 18, 24, and 28 April
2017. We also show the stacked data for the five epochs where no individual
detection could be obtained, which indicates the presence of some level of
quiescent radio emission.  The spectral behavior of the radio emission from
Proxima shows evident changes with time.  This is particularly conspicuous for
the long burst (around 28 April 2017), whose strong total flux density and high
fraction of circular polarization ($|V| / I\gtrsim$80\%) are seen only at
frequencies $\lesssim$2.0 GHz.  A fit to a power-law with frequency for the
data on 28 April 2017 implies a very steep and negative spectral index at
frequencies below $\simeq$2.0 GHz ($\alpha \lesssim -7.0; S_\nu \propto
\nu^\alpha$).  The behavior of this burst appears to be similar in timescale
variability, very steep (negative) spectral index, and degree of circular
polarization to the emission observed at 1.4 GHz in a previous two-day ATCA
campaign \citep{sle+03} in May 2000.

We folded our results to the orbital period of Proxima b ($P_{\rm orb} =
11.186$ days, \citealt{ang+16}) to investigate whether the long-term
variability of the radio emission could be related to the orbital motion of the
planet.  We estimated the times that correspond to the conjunctions of the
planet (i.e.,  when the phase is $\phi = 0$) to obtain the absolute orbital
phase of Proxima b.  We followed a Monte Carlo approach similar to that applied
by \citet{ang+16}, which consists in obtaining a  Markov chain Monte Carlo of
the distribution of orbital parameters derived from the radial velocity
solutions, and then propagating the conjunction prediction for each Markov
chain Monte Carlo step to a number of integer times the orbital period, $P_{\rm
orb}$.  We used all available radial velocities to date (from
\citealt{ang+16}), as well as  50 new HARPS observations obtained in 2017
within the Red Dots campaign, publicly available at the ESO's archive.  This
procedure generated a distribution for the conjunctions that automatically
incorporated the uncertainties in the estimated parameters.  During our
observing campaign, a conjunction happened on JD 2457864.46$^{+0.90}_{-1.09}$
(corresponding to 22:57 UTC on the April 20th, 2017), where the uncertainty
range of [+21.7, -26.1] h corresponds to the 90\% confidence interval.

We show in Fig.~\ref{fig:I_V_vs_orbital_phase} the values of Stokes I and V for
the  low-frequency (1.62 GHz) band data as a function of the  orbital phase of
Proxima b.  The plots evidence two broad emission peaks in both Stokes I and V.
As a metric for the peak of emission, we used the centroid of each broad
emitting region, $\phi_{\rm C}  = \sum_i\, \phi_i\,I_i / \sum_i\,I_i$, where
$\phi_i$ and $I_i$ correspond to the phase and Stokes I, respectively, of each
data point.  The resulting centroids correspond to orbital phases  $\phi_{\rm
C1} = 0.36 \pm 0.01 $ and $\phi_{\rm C2} = 0.81 \pm 0.01$, separated by about
half an orbital period of Proxima b,  when the planet was near the positions of
the quadratures (i.e., when the planet presents the largest angular separation
from the star as seen from Earth).  Since we are interested in the analysis of
the quiescent emission, we excluded the last three data points of the
long-lasting burst in the calculation of the centroids, which is analyzed
separately (see Sect. \ref{sec:two-maxima}).  If we include the flux density
measurements for the long-lasting burst, the second centroid would have a phase
of $\phi_{\rm C2} = 0.73$.  As a metric for the uncertainty in the peak of
emission, we used the rms width of each region, $w = 2\,s$, where $s^2 = \sum_i
(\phi_i - \phi_{\rm C})^2 \, I_i / \sum_i\, I_i $, which yielded values $w_1 =
w_2 = 0.18$, and are shown as hatched areas in
Fig.~\ref{fig:I_V_vs_orbital_phase}.

\subsection{Short-term radio variability} \label{sec:short-duration}

The emission from the Proxima system also displayed intra-day variability, as
shown in Fig.  \ref{fig:flux_evolution}e-h, where we present the observations
averaged over 20 s intervals.  On 2017 April 24, there are two strong, short
duration flares, detected only in the low-frequency band (1.62 GHz), with peaks
at 07:45:10 UT (24.323 Apr) and 14:06:30 UT (24.5875 Apr). Their peak flux
densities ($\sim$25 and $\sim$45 mJy) correspond to about 100 and 200 times the
average flux density value in the rest of the observing session. A possibly
similar short-duration, strongly polarized flare was detected at 1.4 GHz on
1991 August 31 \citep{lim+96}, but apparently caught only at its peak ($\sim$20
mJy) and/or in its decaying phase.

The duration of these flares (estimated as the time where the flux density
exceeds $\sim$2 mJy) is about 4 min in each event, showing a main and a
secondary peak separated by about 2 min, and  hints of substructure at shorter
temporal scales, of about 40 sec.  The two short-duration flares are also
evident in the low-frequency V data (Fig.  \ref{fig:flux_evolution}), and show
a very high degree of circular polarization ($|V|/I$ = 80\%-100\%). On the
contrary, the high frequency band centered on 2.52 GHz does not show any
evidence for those short-duration flares.  

\section{Discussion} \label{sec:discussion}

\subsection{The nature of the observed centimetric radio emission from Proxima Cen} 
\label{sec:periodicity}

\noindent 
The observed high values of brightness temperature and degree of polarization of the
emission for most of the observing epochs require not only a non-thermal origin for the
radio emission, but that the mechanism powering the emission must be a coherent one.
There are two types of coherent mechanisms that can possibly
account for the observed radio emission: plasma emission, as observed in the corona of
some dMe stars  (e.g., \citealt{ste+01}) and ECM emission, as observed in, e.g., the case of
the Jupiter-Io interaction \citep{zar07}.  

\subsubsection{Coherent plasma emission from Proxima Cen} 
\label{sec:plasma-emission}

\noindent Coherent plasma emission is generated by the injection of impulsively heated
plasma with kinetic temperature $T_1 \sim 10^8$ K (hot component) into an ambient plasma
with kinetic temperature $T \sim 10^6$ K (cold component), which causes electron density
oscillations (Langmuir waves) that carry the free energy needed for plasma emission.
For plasma emission to efficiently amplify the radiation, the plasma frequency must be
larger than the gyrofrequency ($\nu_{\rm p} > \nu_g$; \citealt{dul85}), where $\nu_{\rm
p} \approx 9000\, n^{1/2}$ Hz, and $n$ is the plasma density, in cm$^{-3}$. In our
case, $\nu_g \approx 1.62$ GHz, which implies densities $n \gtrsim 3.3\times10^{10}$
cm$^{-3}$ for plasma emission to be efficient. \cite{fuh+11} find plasma density values
$n \sim 5 \times10^{10}$ cm$^{-3}$, which are marginally compatible with the plasma emission
mechanism.

Coherent plasma emission can potentially result in very high brightness
temperatures, $T_b$.  We followed the prescriptions in \citet{ste+01} to
calculate the range of brightness temperatures arising from the coherent plasma
emission mechanism. We find that these can be as large as $T_b \simeq 10^{10}$
K and $\simeq 2.4 \times10^{11}$ K in the fundamental and the second harmonic,
respectively (see Appendix \ref{app:Tb-plasma} for details).  Since the
observed flux densities from Proxima are in the range  from $\sim$174\,$\mu$Jy
up to $\sim$5.0 mJy, the corresponding brightness temperatures are $T_{\rm b}
\gtrsim (1.0 - 31)\times10^{10}\,[\Delta l/ (0.1 R_\ast)]^{-2}$ K.  Fundamental
plasma emission is unlikely to account for the flux density enhancements seen
around 0.36 and 0.81 in phase during our observing campaign. (We note, though,
that \cite{fuh+11} obtained a loop length $\Delta l \sim
8.6^{+3.8}_{-2.9}\times10^9$ cm for a flare of Proxima, in March 2009. In this
case,  the brightness temperature estimates drop to $T_b \sim 10^9$ K, which
would be compatible with fundamental plasma emission, or even an incoherent
emission mechanism, such as (gyro)synchrotron. Thus, while the high circular
polarization and sharp spectral cutoff of the flares are clear indicators of
emission via the ECM mechanism, the estimates of $T_b$ are highly uncertain and
a contribution from (gyro)synchrotron to the non-flaring emission cannot be
ruled out.  Second harmonic plasma emission can reach higher temperatures, and
can more easily account for the observed flux densities. However, the high
degree of polarization observed through our observing campaign is hard to
reconcile with second harmonic plasma emission. While we cannot rule out that
plasma emission has a contribution to the relatively quiescent level of radio
emission observed from Proxima, it  has difficulties in  explaining the
observed characteristics at the times of enhanced emission. For example, while
second harmonic plasma emission can account for the observed brightness
temperatures, the high degree of polarization observed through our observing
campaign does not favor this harmonic emission (for example, second harmonic
emission  in the Sun has shown polarization levels up to about 20\%, much less
than observed in Proxima).

\subsubsection{Electron-cyclotron maser emission from star-planet interaction in Proxima
Cen} 
\label{sec:spi-emission}

The other coherent mechanism capable of producing significant radio emission is
the electron-cyclotron maser emission (ECM) mechanism, which also yields
amplified, highly-polarized radiation.  In the case of star-planet (or
planet-satellite, as in Jupiter-Io) interaction, the friction of the planet
with the magnetic field of the star  generates an unstable population of
electrons that gives rise to significant coherent radio emission. This emission
is constrained within an anisotropic, thin hollow-cone, whose axis coincides
with the local magnetic field vector (\citealt{wu+79,mel+82}), and is visible
only when the walls of this cone are aligned with the observer's line of sight.
The ECM mechanism amplifies mainly one of the two magneto-ionic modes (with
opposite senses of circular polarization) of the electromagnetic wave
propagating within the magnetized plasma \citep{sha+84,mel+84}, which explains
the high degree of circular polarization of the observed radio emission.  

The physical conditions of the region where the ECM efficiently takes place,
namely the ambient plasma density and the strength of the local magnetic field,
define what is the dominant magneto-ionic mode amplified.  The helicity of the
electrons moving within the stellar magnetosphere univocally define the
circular polarization sign of each mode.  Hence, regardless of the amplified
magneto ionic mode, the ECM arising from the two opposites magnetic hemispheres
will be detected as circularly polarized radiation having opposite senses of
polarization.  As an example, the circular polarization sense of the ECM
arising from the early-type magnetic stars carried out clear information
regarding the stellar hemisphere where the ECM originates \citep{let+16}.

In the case of Proxima, the observed radio emission takes place at the expected
ECM frequency for the stellar magnetic field intensity of $\sim$600 Gauss.
The long-term radio emission from Proxima also displays brighter flux
densities, stronger variations, and a higher fraction of circular polarization
in the low-frequency band (centered at 1.62 GHz), compared to the radio
emission in the high-frequency band (centered at 2.52 GHz), in agreement with
expectations from ECM emission due to star-planet interaction.  We  notice that
our observed 1.62 GHz Stokes I flux density, which ranges from $\sim$174
$\mu$Jy up to $\sim$5.0 mJy during our monitoring campaign of Proxima Cen,
broadly agrees with the theoretical flux densities calculated by \citet{tur+18}
for Proxima Cen, who quote values from $\sim$10 $\mu$Jy up to the mJy level.
We also show in Appendix \ref{app:spi-radio} that theoretical estimates of the
Poynting flux arising from star-planet interaction range from as little as
1.4$\times 10^{20}$ erg s$^{-1}$ to as much as 4.4$\times 10^{23}$  erg
s$^{-1}$. These theoretical estimates are broadly consistent  with the Poynting
fluxes inferred from our observations (see Figs.  \ref{fig:Poynt-open}  and
\ref{fig:Poynt-closed}).

\subsubsection{Two maxima of emission per orbital period of Proxima b}
\label{sec:two-maxima}

\begin{figure}[thbp]
  \centerline{\includegraphics[width=\linewidth]{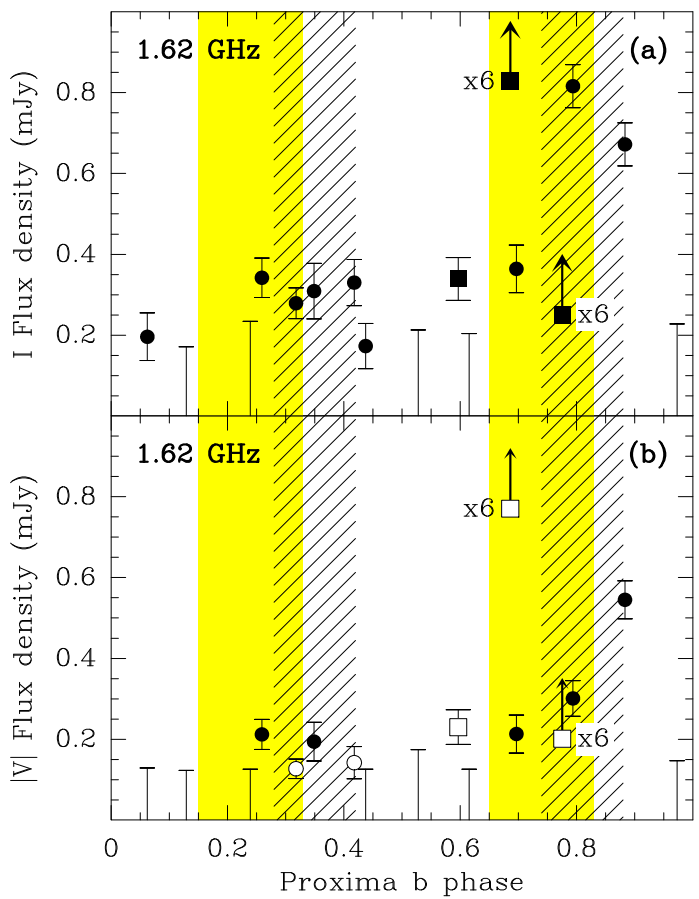}}
  \caption{1.62 GHz flux density light curve of Proxima, folded to the orbital period of
    the planet Proxima b, covering $\sim$1.6 orbital periods.  Variation of the total
    flux density (Stokes I) as a function of the  orbital phase of Proxima b (P$_{\rm
    orb}$=11.186~days; \citealt{ang+16}).  The conjunction time, corresponding to the
    value 0.0 of the orbital phase, is determined to be 20 April 2017, 22:57 UTC, with a
    90\% confidence interval of $[+21.7,-26.1]$~h.  The last three days of observations
    (April 27, 28, and 29; square symbols) correspond to a significantly brighter
    emission burst, and have been analyzed separately.  The 1.62 GHz flux density values are
    higher in two well-defined regions of the orbital cycle of Proxima b.  The centroids
    of these regions correspond to orbital phases of 0.36 and 0.81 (excluding the data
    from the brighter burst), and the hatched regions represent the rms width of the
    maxima of emission.  Yellow-shaded areas correspond to the 90\% confidence interval
    around the 1$^{\rm st}$ and 2$^{\rm nd}$ quadratures.  Filled and open symbols in
    panel \textbf{b} correspond to positive and negative values of Stokes V,
    respectively.  Error bars correspond to 1-$\sigma$ uncertainties.  }
  \label{fig:I_V_vs_orbital_phase} \end{figure}

\noindent Our data indicate the existence of two maxima of emission per orbital
cycle (Fig. \ref{fig:I_V_vs_orbital_phase}).  Two broad emission peaks per
orbital period is the expected behavior for the emission from star-planet
magnetic interaction of the same sort as the Jupiter-Io interaction, which
gives rise to double-peaked auroral radio emission from Jupiter per orbital
cycle of Io.  We notice that any apparent periodicity in our data is unrelated
to the rotation of the Proxima Centauri star, which  has a rotational period
$P_{\rm rot} \approx 83.5$ days \citep{ben+98}.

The maxima of radio emission fall near the quadratures of the planet Proxima b,
also in analogy with the Jupiter-Io interaction where the maxima of radio
emission happen around the quadrature positions \citep{mar+17}.  However, since
our radio observations span only 1.6 orbital periods, we need to assess the
significance of these possible periodic enhancements.  To this end, we estimate
the likelihood that the observed pattern of radio emission from Proxima Cen has
two emission peaks per orbital period, and that these peaks align well with the
known physical periodicity just by mere chance.  The standard way of estimating
this likelihood if the pattern was sinusoidal with a well defined amplitude,
would be by means of a Lomb-Scargle periodogram.  However, the radio emission
does not vary sinusoidally in our case, so we had to revert to a different
method.  Namely, we used the minimum string length (MSL) method that, in
contrast to other methods to build periodograms, is suitable for all sorts of
light curves (single-peaked or multi-peaked, sinusoidal or non-sinusoidal,
etc.), and does not require choosing any parameters.  The MSL method rests on
the fact that the length of a line joining all the points sorted in phase will
be small when a correct period is used to derive the phases.  In this type of
periodogram, potential periods appear as minima in string length plots.
However, while the MSL method and other similar techniques are well suited to
analyze periodic signals with smooth behavior, they are not aimed at analyzing
periodic signals that can have huge excursions in flux from one cycle to the
next.  Therefore, we excluded the data corresponding to the huge flare in the
last three days of our observing campaign (square symbols in
Fig.~\ref{fig:I_V_vs_orbital_phase}).  In this way, we smoothed out the large
flux density variations observed during our campaign. Therefore, we used 14
flux density measurements out of the 17 available. 

We therefore computed the String Length periodograms \citep{dwo83} of 1000
random simulated observing runs (by shuffling the measured flux densities each
time), and compared the simulated minimum string length (MSL) of the light
curves--folded at 11.2 $\pm$0.8 days--with the minimum string length of the
observed flux densities within that interval of periods.  The above quoted
value of $\pm 0.8$ around 11.2 days corresponds to a rough estimate of the
uncertainty with which one might determine the value of the period, using  a
dataset with noise levels and time span such as ours.  The minimum string
length in the random tests was equal to or smaller than the minimum string
length of the observations in 51 out of the 1000 simulated runs.  The
false-alarm probability of the observed configuration of our radio data is thus
at most $p \simeq (51/1000) = 0.051$, since from a visual inspection of those
51 cases, it turns out that about half of the randomly generated light curves
had two emission peaks per orbital cycle.  Hence, the false-alarm probability
of the observed configuration of our radio data is of only $p \simeq (26/1000)
= 0.026$, and the probability that it did not happen by chance is  $1 - p
\simeq 0.97$.  This estimate is a very high value, indicating that it is highly
unlikely that the observed data configuration  happened by mere chance.  If we
use the Stokes V measurements, instead of the Stokes I ones, we obtain similar
results. We also note that the polarized emission also peaks close to the
quadratures, as observed also in the Jupiter-Io system.  

In summary, the observed properties favor the ECM mechanism  over plasma
emission as the coherent mechanism responsible for the radio emission from
Proxima. In particular, the peaked radio emission close to the quadratures is
naturally expected by the ECM mechanism via star-planet interaction, but is
hard to reconcile with the plasma emission mechanism.  In addition, the
observed two broad emission peaks per orbital period, which happen around the
quadrature positions of Proxima b, are unlikely to have occured by mere chance,
and is in analogy with the radio behaviour observed in the Jupiter-Io system.
We therefore suggest that the  ECM instability arising from star-planet
interaction could be the main physical mechanism responsible for the observed
radio emission from Proxima.  In this case, the Proxima/Proxima-b system would
be an analog to the Jupiter/Io system, and a detailed geometrical modeling
(Leto et al. in preparation) is able to explain the observed temporal pattern
and the reversal of the  circular polarization sign of the emission via this
ECM mechanism.  In the case of ECM triggered by star-planet interaction, the
orientation of the magnetic field vector of the stellar magnetosphere
univocally defines the sign of the circularly polarized ECM emission.  Hence,
besides the planet position, also the stellar magnetosphere orientation has a
crucial role for the capability of detecting planet induced ECM emission.  In
fact, two different ECM coherent emissions, occurring close to the quadrature
positions of Proxima b, might arise from opposite hemispheres of Proxima, the
corresponding coherent emissions will be characterized by circular polarization
of opposite signs.  This is the case of the observed reversal in the
polarization sign of the pulses of ECM coherent emission in the ultra-cool
dwarf TVLM513 \citep{hal+07}, whose ECM emission behavior was suggested as an
indirect hint of star-planet interaction \citep{let+17}. 

\subsection{Flaring activity of Proxima} \label{sec:flares}

The two short-flares on 24 April happened 3.3705$\pm$0.0001 days (flare F1) and
3.6313$\pm$0.0001 days (flare F2) after JD 2457864.4563, the estimated time of
the nearest conjunction of the planet Proxima b (Sect. 3.1), that we adopt as
phase reference ($\phi_0=0$). Therefore, the orbital phases  of the two short
flares relative to this reference are $\phi_{\rm F1} -\phi_0 =
0.30131\pm0.00005$ and $\phi_{\rm F2} -\phi_0 = 0.32463\pm0.00005$, where
uncertainties are calculated by error propagation of the uncertainty in the
timing of the peaks of the flares ($\pm$10 s) and in the orbital period of the
planet ($\pm$0.0002 d; Anglada-Escud\'e et al. 2016). Absolute phases,
necessary for a proper comparison with distant events, are largely dominated by
the uncertainty in the calculation of the time of the conjunction adopted to
set the reference phase $\phi_0$. The 90\% confidence interval of this
calculation is [$-$1.09 d,+0.90 d]. Therefore, the absolute orbital phases of
the short flares were $\phi_{\rm F1} = 0.30^{+0.08}_{-0.10}$ and $\phi_{\rm F2}
= 0.32^{+0.08}_{-0.10}$.  Both short-duration flares are consistent with
happening close to the first quadrature of the planet ($\phi_{\rm Q1}=0.25$),
within the uncertainties. This fact, together with the high degree of circular
polarization, suggests that the short-duration flares in Proxima Cen could be
related with a given star-planet orbital position. Indeed, since two flares
were detected during the 10 h on-source of the April 24 session, if these
short-duration flares were randomly distributed in orbital phase, we should
have detected $>1.7$ additional flares of similar intensity (90\% confidence
lower limit assuming small-number Poisson statistics; \citealt{geh86}) during
the 32 h of total on-source time of the remaining 16 sessions.  (We excluded
the April 15 epoch, for which we could not obtain an image in the low-frequency
band.) However, we see no evidence in our data of any other flare of similar
intensity.  If, on the contrary, the occurrence of the short-duration flares
were associated with some specific geometrical configuration (for example, near
a quadrature of the planet, such as on April 24) then the appropriate dates
would be much more restricted, and the expected number of detected flares would
consequently be much smaller, and consistent with the non-detection of
additional short-duration flaring activity. 

The long-lasting burst at the end of our campaign might suggest that the nature
of its radio emission is different from the rest of our data. However, we note
that the peak of the burst happens approximately on 28.8$\pm$0.5 April 2017
(see Fig. 2 and Table 1), or $7.8\pm0.5$ d after the conjunction of reference
where $\phi=0$ is assumed. Taking into account the uncertainties in the timing
of the peak, in the orbital period, and in the reference phase, the absolute
orbital phase obtained is $\phi = 0.70^{+0.09}_{-0.11}$, which is close to the
second quadrature ($\phi_{\rm Q2}$=0.75).  We note that this long burst of
emission shows brighter flux densities, stronger variations and a higher
fraction of circular polarization at frequencies $\leq 2.0$ GHz, and takes
place at the expected electron-cyclotron frequency for the stellar magnetic
field intensity of $\sim$600 Gauss.  In particular,  the emission on 28 April
2017 (Fig.  \ref{fig:Stokes-I-vs-freq}) shows a very steep spectral index
($\alpha \lesssim -7.0; S_\nu \propto \nu^\alpha$) at frequencies below
$\simeq$2.0 GHz, indicative of non-thermal emission.  This emission seems to
switch off abruptly above a frequency of $\sim 2.0$ GHz. The degree of circular
polarization is very large in the low-frequency band ($|V|/I\gtrsim$80\%),
implying a coherent process.  Thus, the characteristics of the long burst are
also consistent with the radio emission being due to the ECM mechanism.

\subsection{Comparison with previous radio observations of Proxima~b} 
\label{sec:flare_comparison}

\cite{sle+03} observed Proxima Cen with ATCA from 14.2 to 15.9 May 2000 and
detected slowly declining radio emission in the 1.38 GHz (22 cm) band. This
emission has a number of similarities with the long-lasting burst at the end of
our observing campaign: it has a very steep and negative spectral index
($\alpha\simeq -12$), a degree of circular polarization close to 100\%, and it
lasted for $\sim$ 2 days or more. The values of the flux density reported by
\citet{sle+03} are of the order of a few mJy, similar to the peak flux density
of the long-lasting burst in our observing campaign. However, these values
correspond to a lower frequency and, given the steep and negative spectral
index of the emission, they would translate into almost 10 times smaller values
at the frequency of 1.62 GHz of our observations, down to the level of what we
call the quiescent emission (Sect.~\ref{sec:images}) that is present in our
observations at epochs far from the quadratures.  Therefore, the radio emission
observed by \cite{sle+03} seems to correspond rather to the decaying stage of a
flare that could be similar to the burst observed at the end of our observing
campaign. Since the flux density decreased as a function of time during the
whole interval of the \cite{sle+03} observations, it must have had a local
maximum before the start of these observations, at an orbital phase $\phi <
0.92^{+0.16}_{-0.12}$. Thus, the orbital phase of the peak of this possible
flare is poorly constrained by their observations.

\cite{lim+96} detected on 31 August 1991 a relatively short-duration (few
minutes to few tens of minutes) flare (peak of $\sim 20$ mJy) towards Proxima
at 20\,cm, with a degree of circular polarization close to 100\%. The flare
happened on 31.12 Aug 1991, corresponding to an orbital phase $\phi =
0.68^{+0.21}_{-0.17}$, which would be consistent with that flare happening at
or around Q2.  This flare could have been similar to the short-duration ones we
detected on 24 April 2017 (Sect.~\ref{sec:short-duration}).

Finally, at much shorter wavelengths, \cite{mac+18} reported a short-duration
($< 1$ min), strong 1.3 mm flare peaking on 2017 March 24 at 08:03 UTC, using
ACA observations.  In contrast to the cm flares, this flare occurred at an
orbital phase of $\phi = 0.53^{+0.09}_{-0.08}$ (close to the planet opposition)
which is, within the uncertainties, incompatible with a quadrature. We note
that since this flare occurred within a few weeks of our ATCA observing
campaign, the relative phasing uncertainty with respect to our flares is very
well constrained to within 0.0004 in phase. We emphasize that while all of the
flares observed at centimeter wavelengths can be explained by being powered by
the ECM mechanism, the flare observed at 1.3 mm occurred at a wavelength where
the ECM, or plasma, coherent mechanisms cannot be powering the observed
emission.

\begin{figure}[htbp]
\centerline{\includegraphics[width=\linewidth]{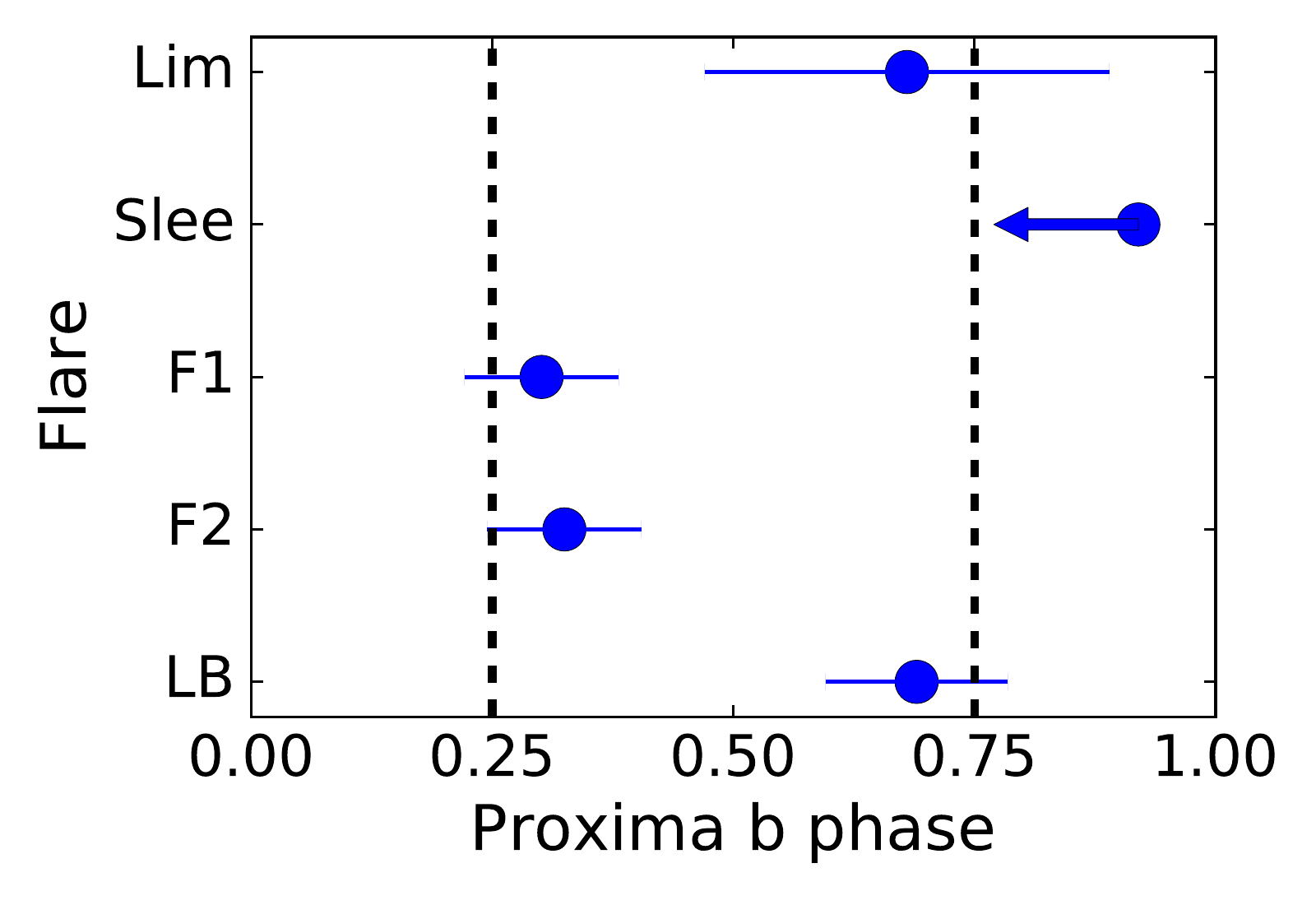}}
\caption{\label{fig:flares} 
    Identified radio flares at centimeter wavelengths (data points) from the Proxima Centauri system against orbital phase of their emission peaks. The dashed vertical lines correspond to the quadrature positions. During our ATCA observations, we observed two short flares, F1 and F2 (Fig.~\ref{fig:flux_evolution} and Sect.~\ref{sec:short-duration}), as well as a long-lasting burst, LB (Fig.~\ref{fig:flux_evolution}). The flares labeled ``Lim'' and `Slee'' correspond to the centimeter-wavelength flares observed back in 1991 and 2000 by \citet{lim+96} and \citet{sle+03} (see Sect.~\ref{sec:flare_comparison}). 
Error bars indicate total uncertainties in the phases of the flares, including the
contributions due to the uncertainty in the time of the flare peak, in the orbital
period and in the absolute phase  reference. The ``Slee'' flare peaked before the start
of those observations and only an upper limit to its phase is known. This is indicated
by the blue arrow. 
}
\end{figure}

In summary, so far five flares have been reported at $\sim$20 cm towards
Proxima Cen: three in our data (the two short flares and the long-lasting
burst), plus the ones reported by \cite{sle+03} and \cite{lim+96} (see
Fig.~\ref{fig:flares}). In four of them (we exclude the Slee et al. flare) the
orbital phase of their peak emission is fairly well constrained and agrees with
a quadrature of the planet Proxima b within the uncertainties (90\% confidence
level) of the orbital parameters. With the current uncertainties, the agreement
between the flare peak and a quadrature is constrained within a phase range of
0.18 [+0.08, -0.10] for each of the two short flares, 0.20 [+0.09, -0.11] for
the long burst and 0.38 [+0.21, -0.17] for the \citet{lim+96} flare.  Since the
probability of the random coincidence of a given flare with a quadrature
(either Q1 or Q2) equals the fraction of the phase-space covered by the
uncertainties of the two quadratures (twice the above values), the probability
of a random coincidence of all the four observed flares with a quadrature
(either Q1 or Q2) is $0.36\times0.36\times0.40\times0.76=0.04$.  This
probability is quite small and hints to a possible relationship between cm
radio emission and the orbital phase of planet Proxima b. Given the additional
properties of the observed emission, this is suggestive of an ECM star-planet
interaction, a possibility that deserves further investigation.  The monitoring
of both the stellar radial velocities and radio emission will better
characterize the occurrence of both quadratures and radio flares, and can
improve our understanding of Proxima Cen system and its magnetic environment.

\section{Summary} \label{sec:summary}

\noindent We observed the Proxima system over 18 consecutive days in April 2017
using the Australia Telescope Compact Array (ATCA) at the frequency band of
1.1-3.1 GHz. Our main findings can be summarized as follows:

\begin{itemize}

  \item We detected radio emission from Proxima for most of the observing
      sessions of our radio monitoring campaign, which spanned $\sim$1.6
      orbital periods of the planet Proxima b and enclosed four quadrature
      positions.  The emission is stronger at the lower frequency band, around
      1.6 GHz, which coincides with the expected electron-cyclotron frequency
      for the star's surface magnetic field intensity of $\sim$600 Gauss, and
      exhibits a large degree of circular polarization, which also reverses its
      sign in the second half of our monitoring campaign. 

\item  The observed radio emission shows a long-term variability with a pattern
    that is consistent with the orbital period of the planet Proxima b around
    the star Proxima.  Namely, the 1.6 GHz radio emission presents  two
    emission enhancements per orbital period of Proxima b, occurring at the
    orbital phase ranges of $0.36\pm0.09$   and $0.81\pm0.09$, i.e., close to
    the quadratures.  The probability that this observed configuration of the
    data happened by chance (the false alarm probability) is $\sim$3\%.

\item Our observations also show a long burst of radio emission that lasted for
    about three days, whose emission peak agrees within the uncertainties with
    the second quadrature of the planet Proxima b. We also detect two
    short-term flares, of a few minutes duration, coincident with the first
    quadrature within the uncertainties. The observed characteristics of the
    radio emission in the long and short-duration flares (frequency around 1.6
    GHz, steep and negative spectral index, peak flux densities of a few to a
    few tens of mJy, degree of polarization close to 100\%, and peak near a
    quadrature) are consistent with being caused by the electron-cyclotron
    mechanism.

\item There is a clustering of the observed centimetric flares around the
    quadrature positions, with all known centimetric flares whose peak has been
    observed (four since 1991) peaking within uncertainties with a quadrature.
    While this does not necessarily imply a precise coincidence between flare
    peaks and quadratures, the probability of all these flares peaking close to
    the quadratures by mere chance is $\sim$4\%), which suggests a relationship
    between this kind of radio emission and the orbital phase of Proxima b. 

\item The ECM emission mechanism accounts well for the observed characteristics
    of the radio emission, and naturally explains the two emission enhancements
    observed per orbital cycle of Proxima b, close to the quadrature positions
    of the planet. Coherent plasma emission, while it may have some
    contribution to the overall observed radio emission, shows characteristics
    that do not match several aspects of our observations, and are hard to
    reconcile with the observed two enhancements per orbital cycle.  The
    observed radio flux densities are also in broad agreement with theoretical
    expectations for ECM emission arising from sub-Alfv\'enic interaction of
    Proxima Cen with its host planet Proxima b. 

\end{itemize}

In summary, the observed 1.6 GHz radio light curve of Proxima Cen shows an
emission pattern that is consistent with the orbital period of the planet
Proxima b around the star Proxima, and its emission peaks happening near the
quadratures.  Furthermore, the properties of the overall observed radio
emission (frequency, large degree of circular polarization, change of the sign
of circular polarization, and observed level of radio emission) are all
consistent with those expected from electron cyclotron-maser emission arising
from sub-Alfv\'enic star-planet interaction.  We therefore favor an
interpretation of our radio observations in terms of interaction between the
star Proxima and its planet Proxima b, which gives rise to the observed radio
emission.  Under this interpretation, the Proxima-Proxima b system may then
represent a scaled-up analog of the observed phenomenology in the Jupiter-Io
system or the Jupiter-Ganymede system, where the planet-moon magnetic
interaction gives rise to electron-cyclotron radio emission in the decametric
spectral region.

The Proxima Cen planetary system, because of its proximity to Earth, is a
particularly valuable target to test the possibility of a detectable
star-planet interaction. Signs of this possible interaction have been
identified in our radio observations thanks to our relatively large monitoring
campaign, as compared to other precedent observations, and to the knowledge of
the planet orbital parameters obtained from optical radial velocity (RV) data.
It is expected that both the radio and the RV data of Proxima Cen will be
significantly improved in the near future, providing a more robust way to
establish and characterize this star-planet interaction, if present.  

Studying the magnetic interaction of other planets around M-dwarf stars (with
intense enough magnetic fields as to emit at decimetric wavelengths), will be
possible with future sensitive radio telescopes, such as the SKA \citep{zar+15}
or its precursors, and may represent a powerful way of detecting and
characterizing exoplanets around stars in the solar neighborhood, and may open
a new field of exoplanet-star plasma interaction studies, thus expanding
magnetospheric and stellar physics.

\begin{acknowledgements} We thank the anonymous referee for the many useful
    comments and suggestions, which improved our paper.  The Australia
    Telescope Compact Array  is part of the Australia Telescope National
    Facility which is funded by the Australian Government for operation as a
    National Facility managed by CSIRO.  M.P.-T, J.F.G., J.L.O., G.A., J.L.G.,
    E.R., A.A., P.A., M.O., M.J.L.-G.,N.M.  and C.R.-L. acknowledge financial
    support from the State Agency for Research of the Spanish MCIU through the
    ``Center of Excellence Severo Ochoa'' award to the Instituto de
    Astrof\'isica de Andaluc\'ia (SEV-2017-0709).  We also acknowledge funding
    support through the following grants: M.P.-T. and A.A.  to
    PGC2018-098915-B-C21 (FEDER/MCIU-AEI); J.F.G., G.A., M.O. to
    AYA2017-84390-C2-1-R (FEDER/MCIU-AEI); J.L.O. to AYA2017-89637-R (MINECO)
    and Junta de Andaluc\'{\i}a 2012-FQM-1776; J.L.G. to AYA2016-80889-P
    (MINECO);
    E.R, to AYA2016-79425-C03-03-P (MINECO) and ESP2017-87676-C05-02-R; 
    P.J.A. and C.R.L. to AYA2016-79425-C03-03-P;  
    M.J.L.-G. to ESP2017-87143-R (MINECO); 
    and Z.M.B. to
CONICYT-FONDECYT/Chile Postdoctorado 3180405.  
\end{acknowledgements}

\begin{appendix}
\section{Brightness temperature of the coherent plasma emission}
\label{app:Tb-plasma}

\noindent We calculated the brightness temperature, $T_b$, for the fundamental
and harmonic of the coherent plasma emission by following the prescriptions in
\citet{ste+01}.  Namely, we assumed a Langmuir wave spectrum between the
wavenumbers $k_{\rm min} = 2\pi\nu_p/v_1$ and $k_{\rm max} =
2\pi\nu_p/(5\,v_{th})$, which takes into account the damping of Langmuir waves
at the thermal background, and implies $T_1 \gtrsim 25\,T$ \citep{ste+01}.
Here, $\nu_p$ is the plasma frequency, which we take equal to 1.62 GHz, $v_1 =
c\, [1 - (m_e c^2/(kT_1 + m_e c^2))^2]^{1/2}$ is the velocity of the hot
electrons, and $v_{th} = (kT/m_e)^{1/2}$ is the thermal velocity of the cold,
ambient electrons. We used a conservative value of 10$^{-5}$ for the fraction
of kinetic energy density of the ambient plasma that goes into the energy
density in Langmuir waves \citep{dul85}.  We used a coronal temperature $T =
2\times10^6$ K, as in Appendix \ref{app:spi-radio}, and estimated the  scale
height, $L_n$, by assuming a hydrostatic density structure for the star, so
that $L_n = k\,T/(\mu\,m_{H}\,g)$, where $\mu$ is the mean atomic weight and
$g$ is the star's gravity. Normalizing to solar values, we obtain $L_n \approx
3.0 \times 10^9\,(T/10^6\, {\rm K})\,(R_\ast/R_\odot)^2\,(M_\ast/M_\odot)^{-1}$
cm. For the Proxima Cen star, we get $L_n \approx 1.0\times10^9$ cm $\approx
0.10\,R_\ast$ ($R_\ast = 0.145\,R_\odot$ is the radius of the Proxima Cen
star).  We then varied  $T_1$  from $5\times10^7$  K to $5\times10^8$ K to
calculate the range of brightness temperatures  for the fundamental, $T_b^{\rm
    f}$,  and the harmonic, $T_b^{\rm h}$ (Eqs. 15 and 16 in \citealt{ste+01}).
    $T_b^{\rm f}$ varies from $6.4\times10^8$ K up to  $1.0\times10^{10}$ K,
    and $T_b^{\rm h}$ varies from $9.9\times10^{10}$ K up to
    $2.4\times10^{11}$ K.  Since our observed Stokes I flux densities, $F$, are
    in the range from $F = 174\,\mu$Jy to $F = 5.0$ mJy, the corresponding
    observed brightness temperatures are $T_{\rm b} \gtrsim (1.0 -
    31)\times10^{10}\,[\Delta l/ (0.1 R_\ast)]^{-2}$ K. Therefore, fundamental
    plasma emission could marginally account for the quiescent level of radio
    emission detected during our observing campaign, but has difficulties in
    accounting for the flux density enhancements seen around the quadratures.  

\section{Radio energetics from star-planet interaction}
\label{app:spi-radio}

\noindent Here, we discuss the feasibility that the radio emission arising from
star-planet interaction in Proxima Cen can be detected in our observations, and
follow the formalism of \citet{ved+20}. 

Theoretical estimates of the Poynting flux due to star-planet interaction in
the sub-Alfv\'enic regime (i.e., when the relative velocity between the stellar
wind flow and the planet, $v_{rel}$, is smaller than the plasma Alfv\'en
velocity, $v_A$) at the location of the planet, are given in, e.g.,
\citet{zar07}, \citet{lan09}, \citet{sau+13}, and \citet{tur+18}.  These
estimates indicate that the total Poynting flux is $S^{\rm th}_{Poynt} =
R_{eff}^2v_{rel}B_{sw}^2 \, \epsilon/2$, where $R_{eff}$ is the effective
radius of the planetary obstacle, $B_{sw}$ is the stellar wind magnetic field
at the location of the planet, and $\epsilon \leq 1$  encapsulates efficiency
and geometric factors related to the nature of the interaction. We can rewrite
the theoretical total Poynting flux as follows:

\begin{equation}
  \label{eq:poynt-th}
  S^{\rm th}_{Poyn} \approx   1.2 \times 10^{21}\,
\left(\frac{R_{\rm eff}}{1.1\,R_\Earth}\right)^2\, 
\left(\frac{v_{rel}}{500\, {\rm km s^{-1}}}\right)\, 
    \left( \frac{B_{\rm sw}}{0.1\,G} \right)^2\, 
    \left( \frac{\epsilon}{0.01}\right)\, 
    {\rm erg \,s^{-1}},
\end{equation}

where we have normalized the effective radius of the obstacle, $R_{\rm eff}$, to the
radius of Proxima b, which is $R_p \approx 1.1 R_\Earth$ \citep{bix+17}.

$S^{\rm th}_{Poyn}$ can be compared with the Poynting flux inferred from the
observed radio emission, $S_{Poyn}^{\rm obs}$.  The total emitted radio power
is $P_{R} = F\,\Omega\, D^2\,\Delta\nu$, where $F$ is the observed radio flux
density, $\Omega$ is the solid angle into which the ECM radio emission is
beamed, $D$ is the distance to the star, and $\Delta\nu$ is the total bandwidth
of the ECM emission.  We assume a typical bandwidth for the ECM emission of
$\Delta\nu = \nu_{\rm g}/2$, where $\nu_g \approx 2.8\, B_\ast$ MHz is the
cyclotron frequency, and $B_\ast$ is the average surface magnetic field
strength of the Proxima Cen star. 
The observationally inferred  Poynting flux is thus $S_{Poyn}^{\rm obs} = P_{R}
/ \epsilon_{\rm rad}$, where the factor $\epsilon_{\rm rad}$ corresponds to the
efficiency in converting the Poynting flux into ECM emission. 
For $D=1.3$ pc, and using an average flux density value at 1.62 GHz of $F
\approx 0.31$ mJy, the power emitted is $P_{R} \approx 4.2\times10^{19} \,
(F/300\, \mu{\rm Jy}) \,(B_\ast/600 {\rm G})\, (\Omega/{\rm 1\,sr})$ erg
s$^{-1}$, and $S_{Poyn}^{\rm obs}$ can then be written as

\begin{equation}
  \label{eq:poynt-obs}
  S^{\rm obs}_{Poyn} \approx 4.2\times10^{21} \, \left(\frac{F}{300\, \mu{\rm Jy}}\right) \,
\left(\frac{B_\ast}{600\,{\rm G}}\right) \,\left(\frac{\Omega}{\rm 1\,sr}\right) \left(\frac{\epsilon_{rad}}{0.01}\right)^{-1}
{\rm erg \,s}^{-1},
\end{equation}

The  efficiencies in the conversion of Poynting flux into ECM emission (the
factor $\epsilon_{rad}$) are estimated to be in the range from about 1\%
\citep{asc90} up to values of 10\% or even higher \citep{kuz11}.  Equations
\ref{eq:poynt-th} and \ref{eq:poynt-obs} show that star-planet interaction can
potentially result in Poynting fluxes large enough that detection of its
centimetric radio emission from Earth is feasible.

We assume that the electrons responsible for the cyclotron emission have
kinetic energies between $E_{k, min}=$10 keV and up to the rest-mass of the
electron, $E_{k, max} = m_e\,c^2 = $ 511 keV.  The speed of the
electrons, $\beta$, depends on the Lorentz $\gamma$ factor as follows: $\beta =
(1 - \gamma^{-2})^{1/2}$, where $\gamma = 1 + E_k/(m_e\,c^2)$. Therefore, the above
range of kinetic energies  correspond to $\beta$ in the range $[0.20, \,
0.87]$.  We make the standard assumption that the electrons emit from within a
cone with half-opening angle $\theta$ and angular width $\Delta\theta$, which
are related to $\beta$ as follows: $\cos \theta \approx \Delta\theta \approx
\beta$ \citep{mel+82}.  For our values of $E_{k, min}$ and $E_{k, max}$, the
beam solid angle subtended by the emission cone  is in the range from 1.20 sr
up to 2.6 sr.

\begin{figure}[h!]
  \centerline{\includegraphics[width=0.9\linewidth]{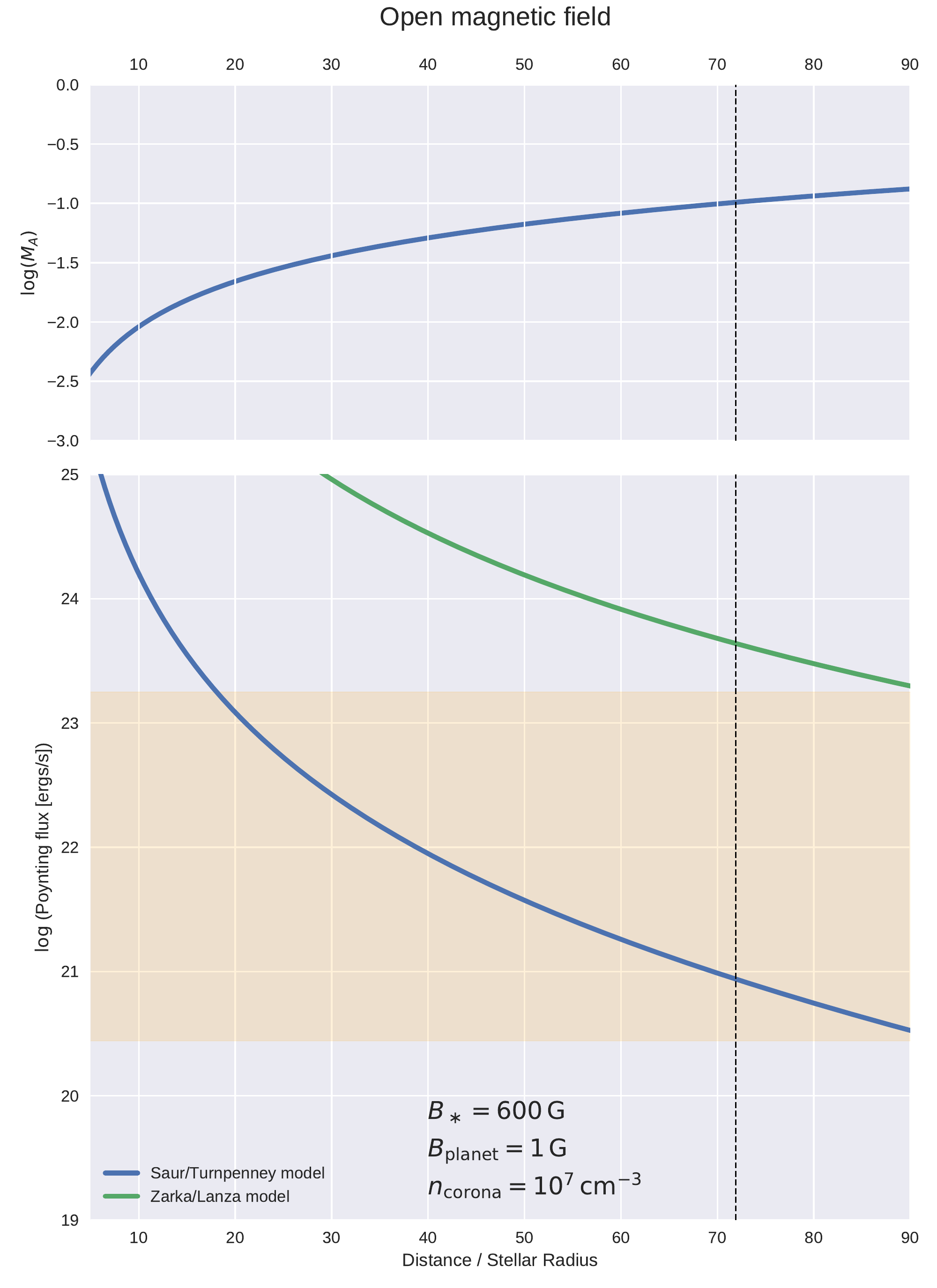}} 
\caption{ 
  Comparison of theoretical expectations and observationally inferred values 
  of the Poynting flux from sub-Alfv\'enic interaction in Proxima, 
  for an open Parker spiral magnetic field geometry, as a function of the
  radial distance to the Proxima Cen star. 
  The upper panel shows the Alfv\'en number, $M_A$. 
  The curves in the lower panel correspond to the theoretical Poynting flux, 
  $S^{\rm th}_{Poyn}$, for  two different 
  models of the interaction: the Saur/Turnpenney model
  (\citealt{sau+13,tur+18}; solid blue line) and the 
  Zarka-Lanza model (\citealt{zar07,lan09}; solid green line). The orange-shaded region corresponds to the range
  of observationally inferred Poynting fluxes 
  $S^{\rm obs}_{Poyn}$, allowed by our observed radio flux densities, and the
  dashed line is drawn at the orbital distance of Proxima b.
}\label{fig:Poynt-open}
\end{figure}

\begin{figure}[h!]
  \centerline{\includegraphics[width=0.9\linewidth]{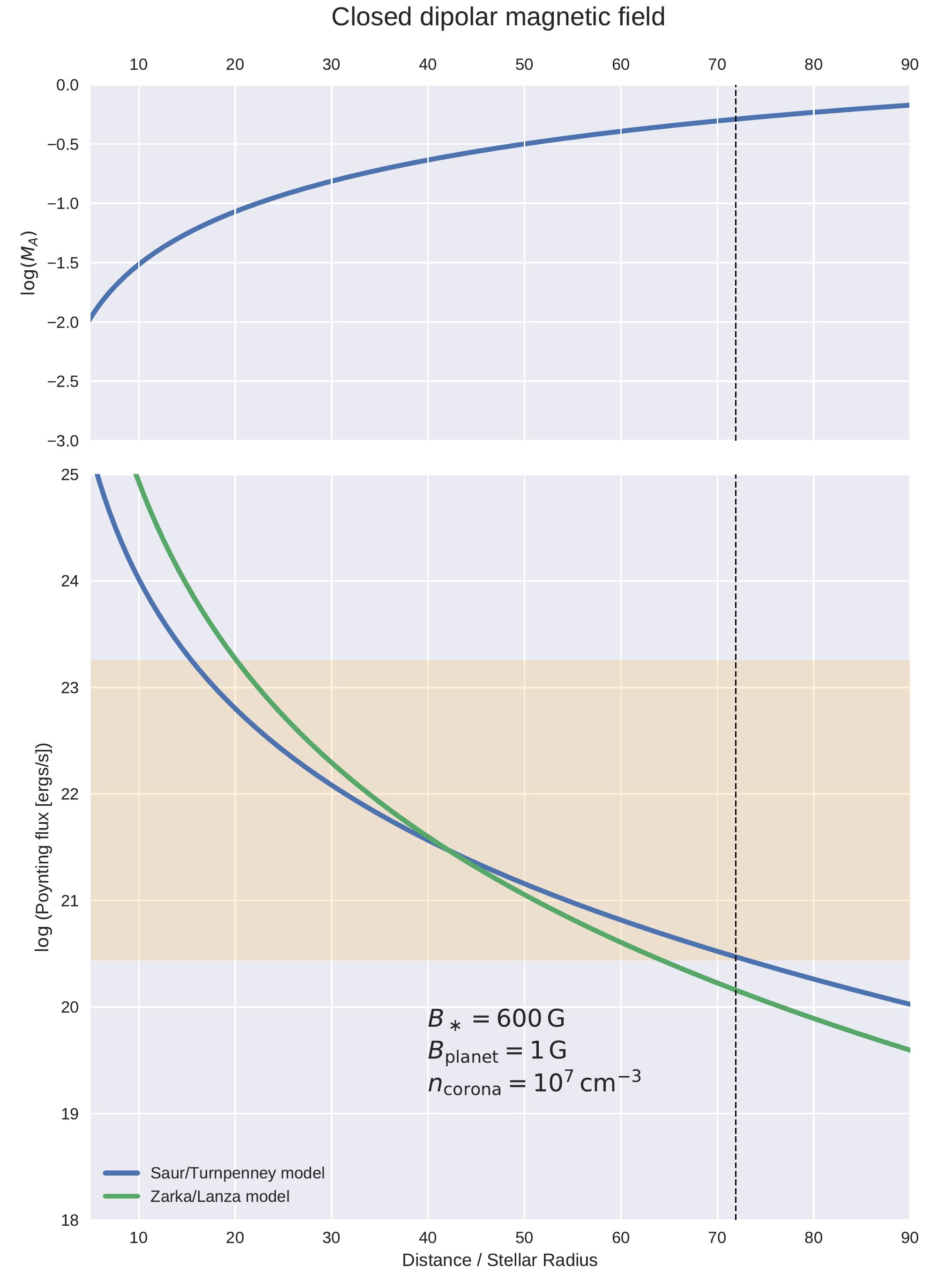}} 
\caption{ 
  Same as in Fig. \ref{fig:Poynt-closed}, but for a closed dipolar geometry. 
}\label{fig:Poynt-closed}
\end{figure}

We discuss two cases of the magnetic field geometry for the star-planet
interaction:   a close-field, dipole geometry,  and an open-field, Parker
spiral geometry, also as in \citet{ved+20}.  For each case, we consider two
models for the efficiency of the interaction.  One model follows the
prescriptions by \citet{sau+13} and \citet{tur+18}, where $\epsilon =
\bar{\alpha}^2\,M_A\,\sin^2\Theta$. Here, $M_A = v_{rel}/v_A$ is the Alfv\'en
number at the planet location,  $\Theta$ is the angle between the stellar wind
magnetic field at the planet and the stellar wind velocity in the frame of the
planet (e.g., \citealt{sau+13,tur+18}), and $\bar{\alpha}$ is the plasma
flow-obstacle interaction strength factor, which for the case of Proxima b is
well approximated by $\bar{\alpha} \simeq 1$ \citep{tur+18}.  We followed the
prescriptions given in the Appendix B of \citet{tur+18} to determine the speed
of the stellar wind, $v_{sw}$, the magnetic field of the wind, $B_{sw}$, and
the angle $\Theta$.  As in \citet{tur+18}, we used an isothermal stellar wind
\citep{par+58}, which is fully parameterized by the sound speed, or
equivalently, the coronal temperature, $T$.  We adopted $T = 2\times10^6$ K
for the coronal temperature of Proxima Cen, which agrees well with the
temperatures inferred from X-ray observations (e.g.,  \citealt{fuh+11}).  
The other model follows \citet{zar07} and \citet{lan09}, where $\epsilon = \eta
/ 2$ and $\eta$ is a geometric factor, which we assume to be $\eta = 1/2$. We
note that, since $\sin \Theta \leq 1$ and usually $M_A \ll 1$, the Zarka-Lanza
model predicts significantly larger Poynting fluxes than the Saur-Turnpenney
model (see Figs. \ref{fig:Poynt-open} and \ref{fig:Poynt-closed}).

We adopted $B_\ast =$ 600 G for the stellar surface magnetic field
\citep{rei+08}, which falls down with radial distance as $r^{-3}$ (closed
dipole geometry) and as $r^{-2}$ (open field geometry).  We obtained the
effective obstacle radius, $R_{eff} (\ge R_p) $,  by balancing the pressure of
the planet's magnetosphere with that of the stellar wind flow. We followed
\citet{lan09} and took $R_{eff}$ to be the distance from the planet at which
the stellar and planetary magnetic fields are equal, i.e., $R_{eff} =
R_p\,(B_p/B_{sw})^{1/3}$, where $B_{sw}$ is the magnetic field of the wind at
the orbital distance of Proxima b, and $B_p$ the planetary magnetic field.
$R_{eff}$ is  further modified by a factor of order unity that depends on the
angle $\Theta_M$ between the magnetic moment of the planet and the stellar
magnetic field \citep{sau+13}, which we set equal to $\Theta_M = 0$ and
$\Theta_M = \pi/2$ for the closed- and open-field cases, respectively. 

We show in Figs. \ref{fig:Poynt-open} and \ref{fig:Poynt-closed} the
theoretically expected and observationally inferred range of values for the
Poynting flux for our adopted nomimal model with $n_{\rm corona} = 10^7\,{\rm
cm}^{-3}, T = 2\times\,10^6$ K and a planetary magnetic field of $B_p = 1$ G.
We obtained the plasma density at the orbital distance of Proxima b by letting
evolve the value of density at the base of the stellar corona, $n_{\rm corona
}$, with radial distance as $r^{-2}$.  Fig. \ref{fig:Poynt-open} corresponds to
a Parker spiral (open) geometry of the magnetic field, while Fig.
\ref{fig:Poynt-closed} is for a dipolar (closed) magnetic field geometry. The
theoretical expectations for the Poynting flux are drawn as solid lines (blue:
Saur/Turnpenney model; green: Zarka/Lanza model), while  observationally
inferred values are drawn as light orange-shaded areas.  The orange-shaded
areas in both figures correspond to the range of observationally inferred
Poynting fluxes,  $S^{\rm obs}_{Poyn}$, for our observed   Stokes I flux
densities (from $F = 174\,\mu$Jy up to $F = 5.0$ mJy), taking into account the
range of beam solid angles of the emission (see above), and the range of the
efficiencies in converting Poynting Flux into radio emission, $\epsilon_{rad}$,
which we took to be from 1\% up to 10\%.  For the nominal values of $B_\ast,
B_p$ and $n_{\rm corona}$, the theoretically expected Poynting fluxes in the
open field case are of $S^{\rm th}_{Poyn}$ of  8.7$\times 10^{20}$ erg s$^{-1}$
and 4.4$\times 10^{23}$  erg s$^{-1}$ for the Saur/Turnpenney model and the
Lanza/Zarka model, respectively (Fig. \ref{fig:Poynt-open}).  In the closed
magnetic field case, the expected values of $S^{\rm th}_{Poyn}$ are  3.0$\times
10^{20}$ erg s$^{-1}$ and 1.4$\times 10^{20}$  erg s$^{-1}$ for the
Saur/Turnpenney model and the Lanza/Zarka model, respectively (Fig.
\ref{fig:Poynt-closed}).

\begin{figure}[h!]
  \centerline{\includegraphics[width=0.9\linewidth]{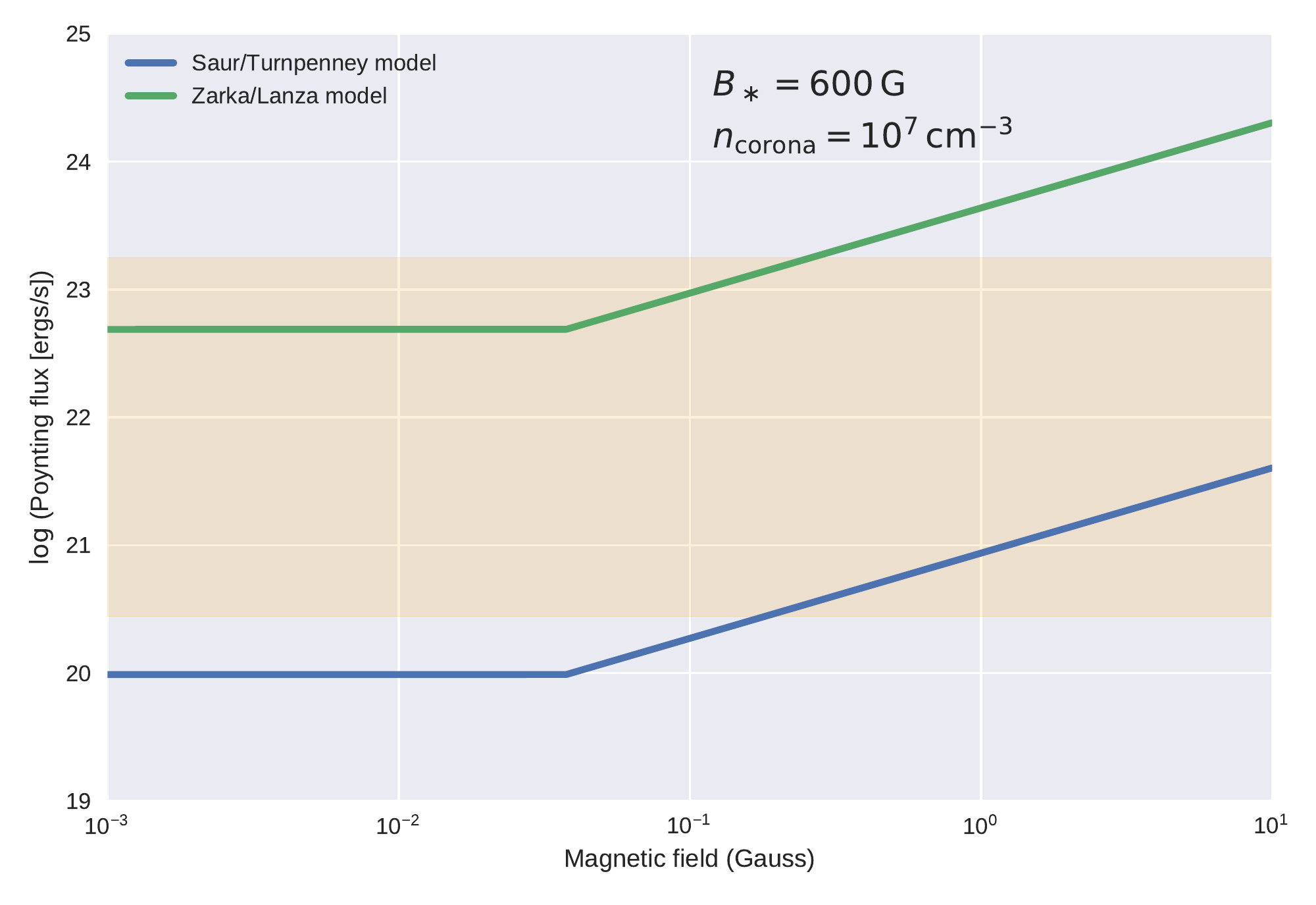}}
    \centerline{\includegraphics[width=0.9\linewidth]{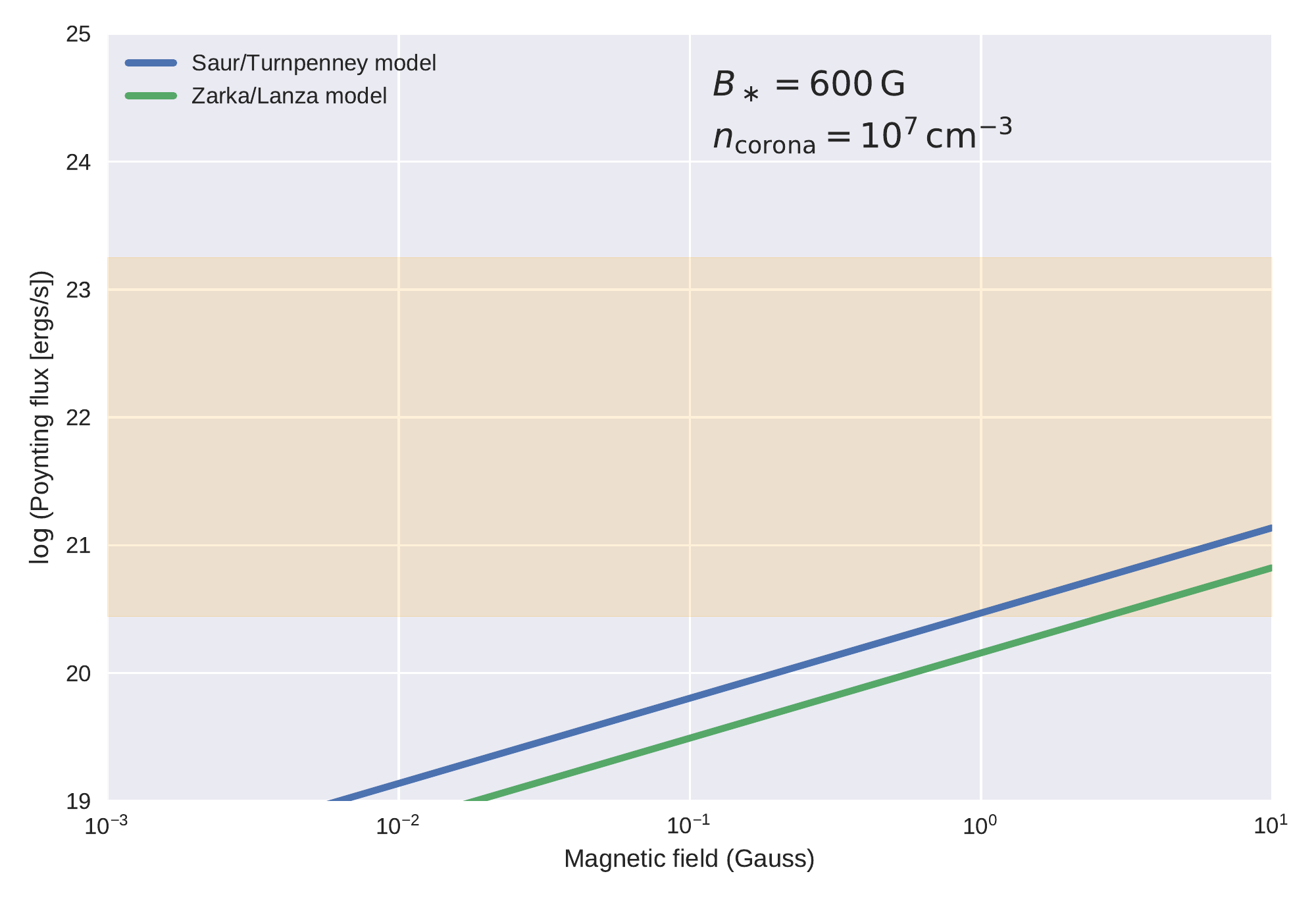}} 
\caption{ 
Comparison of theoretical expectations and observationally inferred values 
of the Poynting flux from sub-Alfv\'enic interaction in Proxima, 
as a function of the magnetic field of the planet Proxima b. 
Top: Open magnetic field; bottom: Closed magnetic field.
}\label{fig:Poynt-Bp}
\end{figure}

\begin{figure}[h!]
  \centerline{\includegraphics[width=0.9\linewidth]{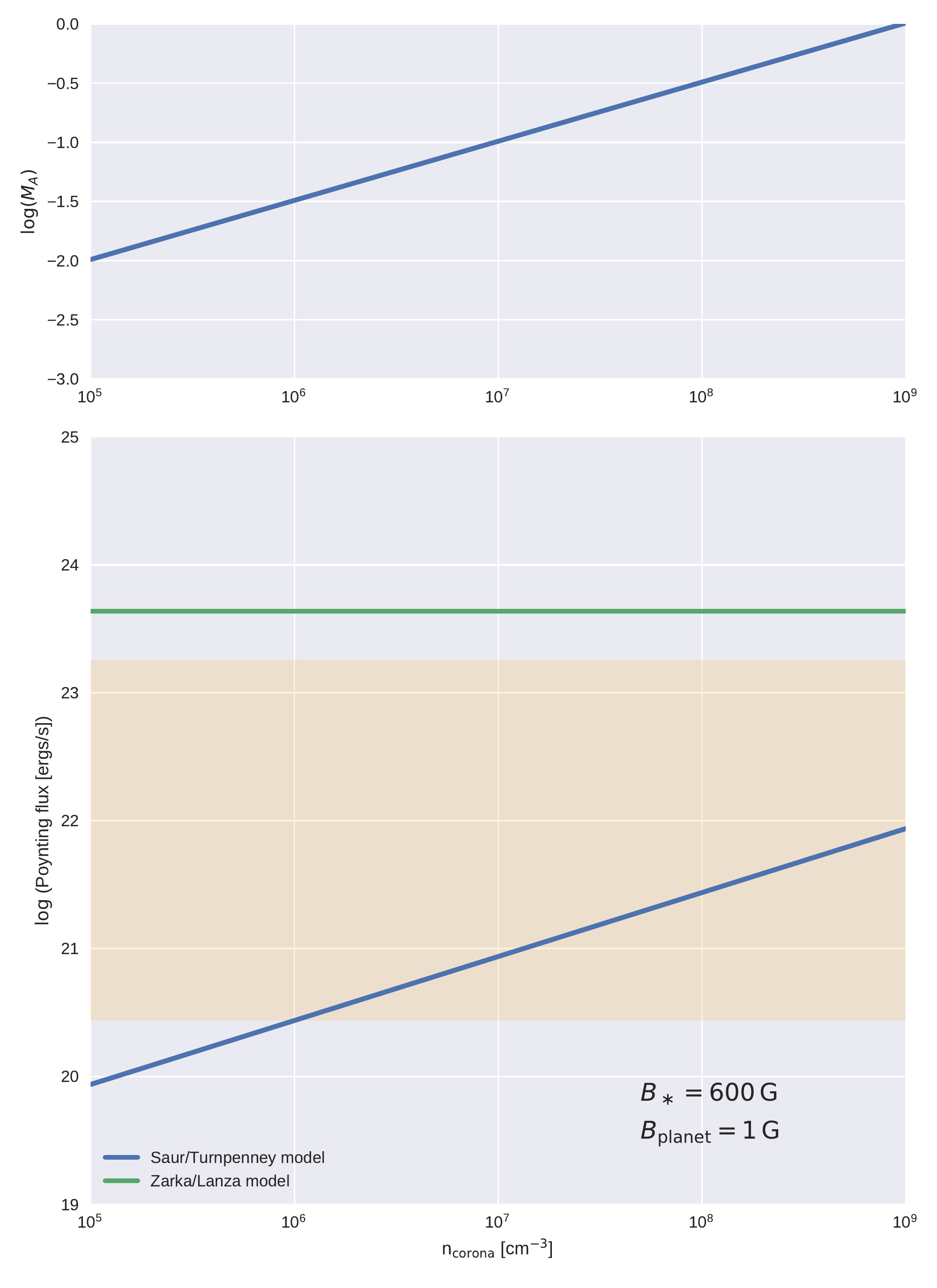}}
\caption{ 
Comparison of theoretical expectations and observationally inferred values 
of the Poynting flux from sub-Alfv\'enic interaction in Proxima, 
as a function of density at the base of the stellar corona, for an open magnetic field geometry.
}\label{fig:Poynt-ne-open}
\end{figure}

\begin{figure}[h!]
    \centerline{\includegraphics[width=0.9\linewidth]{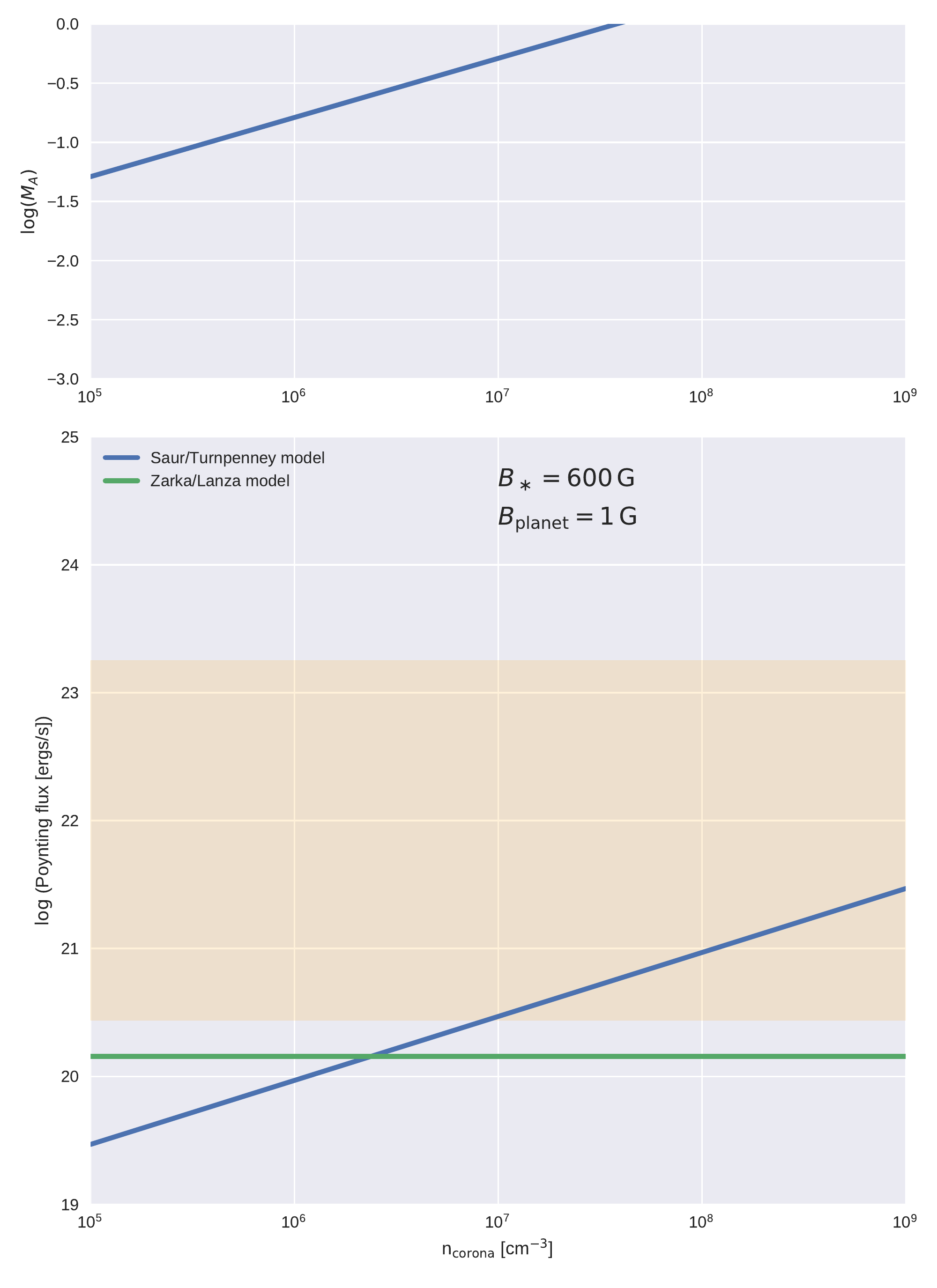}} 
\caption{ 
Same as in Fig. \ref{fig:Poynt-ne-open}, but for a closed magnetic field. The regime stops being sub-Alfv\'enic at the orbital distance of Proxima b for $n_{\rm corona} \approx 3.8\times10^{7}$ cm$^{-3}$, corresponding to a density of about 
7000 cm$^{-3}$ at the orbital position of Proxima b.
}\label{fig:Poynt-ne-close}
\end{figure}

Figure \ref{fig:Poynt-Bp} shows the dependence of the Poynting flux (computed
at the orbital distance of Proxima b) with the magnetic field of the planet,
$B_p$, for both magnetic field geometries.  We note that, since the magnetic
field of the wind at the position of Proxima b is significantly larger in the
open-field case than in the closed-field one, the Poynting flux is
correspondingly larger.  In the open-field case, $S^{\rm th}_{Poyn}$ is
constant for planetary magnetic fields  below $\sim$40 mG because for smaller
values of $B_p$, the effective radius of the obstacle, $R_{eff}$, equals the
planet radius $R_p$.  We show in Figs.  \ref{fig:Poynt-ne-open} and
\ref{fig:Poynt-ne-close} the dependence of the Alfv\'en number and Poynting
flux with the density at the orbital distance of Proxima b, $n_p$. Since
$r_{\rm orb}/R_\ast = 71.9$, the plasma density at the orbital position of
Proxima b is related with the density at the base of the corona as follows:
$n_p = n_{\rm corona}\,({r_{\rm orb}/R_\ast})^{-2} \approx
1.9\times10^{-4}\,n_{\rm corona}$ For large densities, the regime becomes
supra-Alfv\'enic, and hence the Poynting fluxes do not apply.  We also note
that the Poynting flux predicted by the Saur/Turnpenney model grows with
density as $n_p^{1/2}$ (for all other parameters fixed), while that predicted
by the Zarka/Lanza model remains constant. This is because $S_{Poyn}^{\rm th}
\propto\, v_A\,M_A^2 \propto\, n_p^{-1/2}\,n_p = n_p^{1/2}$ in the former model
\citep{sau+13}, while $S_{Poyn}^{\rm th} \propto\, v_A\,M_A \propto\,
n_p^{-1/2}\, n_p^{1/2} = $ constant in the latter.

Those figures illustrate that star-planet interaction  between the Proxima star
and its planet Proxima b is capable of yielding Poynting fluxes that are
broadly consistent with the observed radio flux densities. 

\end{appendix}

\end{document}